\documentclass[iop]{emulateapj}

\usepackage{color}
\usepackage{natbib}
\usepackage{amsmath}
\usepackage{gensymb}
\usepackage{appendix}
\usepackage{array}
\shorttitle{Mass-Radius-Period Distribution}
\shortauthors{Neil \& Rogers}

\begin{document}

\title{A Joint Mass-Radius-Period Distribution of Exoplanets}

\author{Andrew R. Neil and Leslie A. Rogers}
\affil{The Department of Astronomy and Astrophysics, University of Chicago,
    Chicago, IL 60637}

\begin{abstract}
The radius-period distribution of exoplanets has been characterized by the \textit{Kepler} survey, and the empirical mass-radius relation by the subset of \textit{Kepler} planets with mass measurements. We combine the two in order to constrain the joint mass-radius-period distribution of \textit{Kepler} transiting planets. We employ hierarchical Bayesian modeling and mixture models to formulate four models with varying complexity and fit these models to the data. We find that the most complex models that treat planets with significant gaseous envelopes, evaporated core planets, and intrinsically rocky planets as three separate populations are preferred by the data and provide the best fit to the observed distribution of \textit{Kepler} planets. We use these models to calculate occurrence rates of planets in different regimes and to predict masses of \textit{Kepler} planets, revealing the model dependent nature of both. When using models with envelope mass loss to calculate $\eta_\oplus$, we find nearly an order of magnitude drop, indicating that many Earth-like planets discovered with \textit{Kepler} may be evaporated cores which do not extrapolate out to higher orbital periods. This work provides a framework for higher-dimensional studies of planet occurrence and for using mixture models to incorporate different theoretical populations of planets. 
\end{abstract}


\section{Introduction}\label{Introduction}

With the advent of large-scale exoplanet surveys, exemplified by NASA's \textit{Kepler} mission \citep[e.g.][]{BoruckiEt2011ApJ}, we have arrived at a much more detailed picture of the demographics of exoplanets. The \textit{Kepler} survey has discovered thousands of transiting exoplanets, and with its well-studied detection efficiency, has characterized the radius-period distribution of small planets to high precision \citep{Burke2015ApJ, ThompsonEt2018ApJS}. Early studies were able to reveal that planets smaller than Neptune are common, with an occurrence rate (number of planets per star) on order unity \citep{HowardEt2012ApJS, FressinEt2013ApJ}. Later studies improved methodology by employing hierarchical Bayesian modeling \citep{Foreman-MackeyEt2014ApJ, Burke2015ApJ} as well as approximate Bayesian computation \citep{HsuEt2018AJ}, avoiding a bias in earlier papers that underestimated the occurrence rate of planets at the threshold of detection. Later studies also benefited from a better characterized detection efficiency \citep{ChristiansenEt2015ApJ,Christiansen2017Tech}, as well as a stellar sample with high-precision radii and masses \citep{PetiguraEt2017AJ}, the latter allowing for the discovery of a gap in the radius distribution of small planets between $1.5-2.0 R_\oplus$ \citep{FultonEt2017ApJ}. Finally, with the release of Data Release 25, the final data release of \textit{Kepler}, we have the best characterization of the radius-period distribution of transiting planets to date \citep{Fulton&Petigura18AJ, MuldersEt2018AJ, HsuEt2019Arxiv}.

While the radius-period plane is the most natural plane to constrain planet occurrence rates for transit surveys, it is but a projection of a high dimensional space where the occurrence of planets can depend on many factors, both those intrinsic to the planet and those relating to the host star that the planet orbits. Planet mass is one of the most fundamental properties of a planet and has been studied statistically through radial velocity (RV) and microlensing surveys \citep{HowardEt2010Science, MayorEt2011Arxiv, SuzukiEt2016ApJ}. With RV follow-up of transiting planets, the mass-radius distribution of planets, commonly characterized by a mass-radius relationship, has been empirically constrained. Since the data is limited by the rate of RV and transit-timing variation (TTV) characterization of small planets, most studies have parametrized the mass-radius relation as a power-law with multiple breaks \citep{Wu&Lithwick2013ApJ,WeissEt2013ApJ,Weiss&Marcy2014ApJL}. Later studies have improved upon the statistical methodology and have accounted for the intrinsic scatter in the mass-radius relation \citep{WolfgangEt2016ApJ,ChenKipping2017ApJ,Neil&Rogers2018ApJ, NingEt2018ApJ}. While the theory of planet interior structure and its manifestation in the mass-radius relation has been studied for over a decade \citep{ValenciaEt2006Icarus, FortneyEt2007aApJ, SeagerEt2007ApJ, RogersEt2011ApJ, Lopez&Fortney2013ApJ, Howe&Burrows2015ApJ, ZengEt2016ApJ, Chen&Rogers2016ApJ}, these efforts have largely been separate from empirically derived mass-radius relations.

The 2D distributions of radius-period, mass-period, and mass-radius have all been characterized using various datasets and statistical techniques. To date, no effort has been made to constrain all three distributions simultaneously and self-consistently. In this paper, we constrain the 3D joint mass-radius-period distribution of exoplanets using the \textit{Kepler} survey, a subset of which have mass measurements. In doing so, we can combine the state-of-the-art data and techniques used to constrain each 2D distribution, and expand the analysis into 3D space. This 3D joint distribution can then be marginalized to obtain each individual 2D distribution, which can be used for such purposes as calculating occurrence rates or predicting individual masses or radii of planets.

This paper is organized as follows. In section \ref{Data} we outline how we arrived at our planet sample and detection efficiency function. Four joint mass-radius-period distribution models with increasing complexity are presented in section \ref{Models}. The fitting process is described in section \ref{Fitting}. The results of our model fitting are described in section \ref{Results}, along with model selection, occurrence rate calculations and individual planet mass and radius predictions. We discuss caveats, how to use these results and future extensions of this work in section \ref{Discussion}, and conclude in section \ref{Conclusion}.

\section{Data}\label{Data}

In order to constrain the joint mass-radius-period distribution, we use the California-\textit{Kepler} Survey (CKS), a subset of transiting planets from \textit{Kepler} with high-resolution spectroscopic follow-up of their host stars \citep{PetiguraEt2017AJ, JohnsonEt2017AJ}, cross-matched with \textit{Gaia} data. This sample has several significant advantages. First, it is by far the largest sample of planets with high-precision radius measurements. The CKS sample was the first sample to uncover the gap in the \textit{Kepler} radius distribution, due to the low median uncertainty in planet radii of $13\%$ \citep{FultonEt2017ApJ}. Cross-matching the sample with \textit{Gaia} parallaxes achieves even lower radius uncertainties (median of 5\%) and allowed a stellar mass dependence in the radius gap to be revealed \citep{Fulton&Petigura18AJ}. Second, there is a subset of planets in the sample that have mass measurements, either through RV or TTV, allowing us to constrain the mass-radius relation. Finally, the \textit{Kepler} survey has a well-defined detection efficiency as a function of period and radius, essential for correcting for the detection bias in our estimate of the true occurrence rate density of planets. 

We follow \citet{Fulton&Petigura18AJ} and apply cuts to the CKS sample to ensure high quality data. In brief, we use the magnitude-limited sample (brighter than $14.2$ \textit{Kepler} magnitude), take out evolved stars with a $T_{\text{eff}}$-dependent cut on $R_{*}$, and further limit the sample to G and K dwarfs ($4700\ \text{K} < T_{\text{eff}} < 6500\ \text{K}$). Additionally, we remove stars that have high dilution based on \textit{Gaia} and imaging data. We remove any planet candidates that have been designated as false-positives by the \textit{Kepler} team, and any planets with grazing transits ($b > 0.9$) due to uncertainty in planet radii at high impact parameters. Lastly, we remove any planets with properties that fall outside our orbital period and radius bounds, which are $0.3 - 100$ days and $0.4 - 30 R_{\oplus}$, respectively (see section \ref{Models} for why these bounds were chosen). The final planet radii we use are taken from \citet{Fulton&Petigura18AJ}, which were calculated using stellar radii derived from both CKS spectroscopy and \textit{Gaia} parallaxes. Our final sample has 1130 planets, with a median radius uncertainty of $4.8 \%$.

On the mass side of things, we use mass measurements where available for planets in the CKS-\textit{Gaia} sample. Given the systematic differences in the density of planets measured with RV vs TTV \citep{JontofHutterEt2014ApJ, Steffen2016MNRAS, Mills2017ApJL}, we limit our mass sample to RV-measured masses only, leaving the analysis including TTV masses to a future work. We discuss the impact of this choice in section \ref{Data Set}. We take RV mass measurements from the NASA Exoplanet Archive\footnote{https://exoplanetarchive.ipac.caltech.edu/}, downloaded on September 25, 2018. The CKS-\textit{Gaia} sample we use in this paper has 53 planets with RV mass measurements.

\begin{figure}
    \centering
    \includegraphics[width=1\columnwidth]{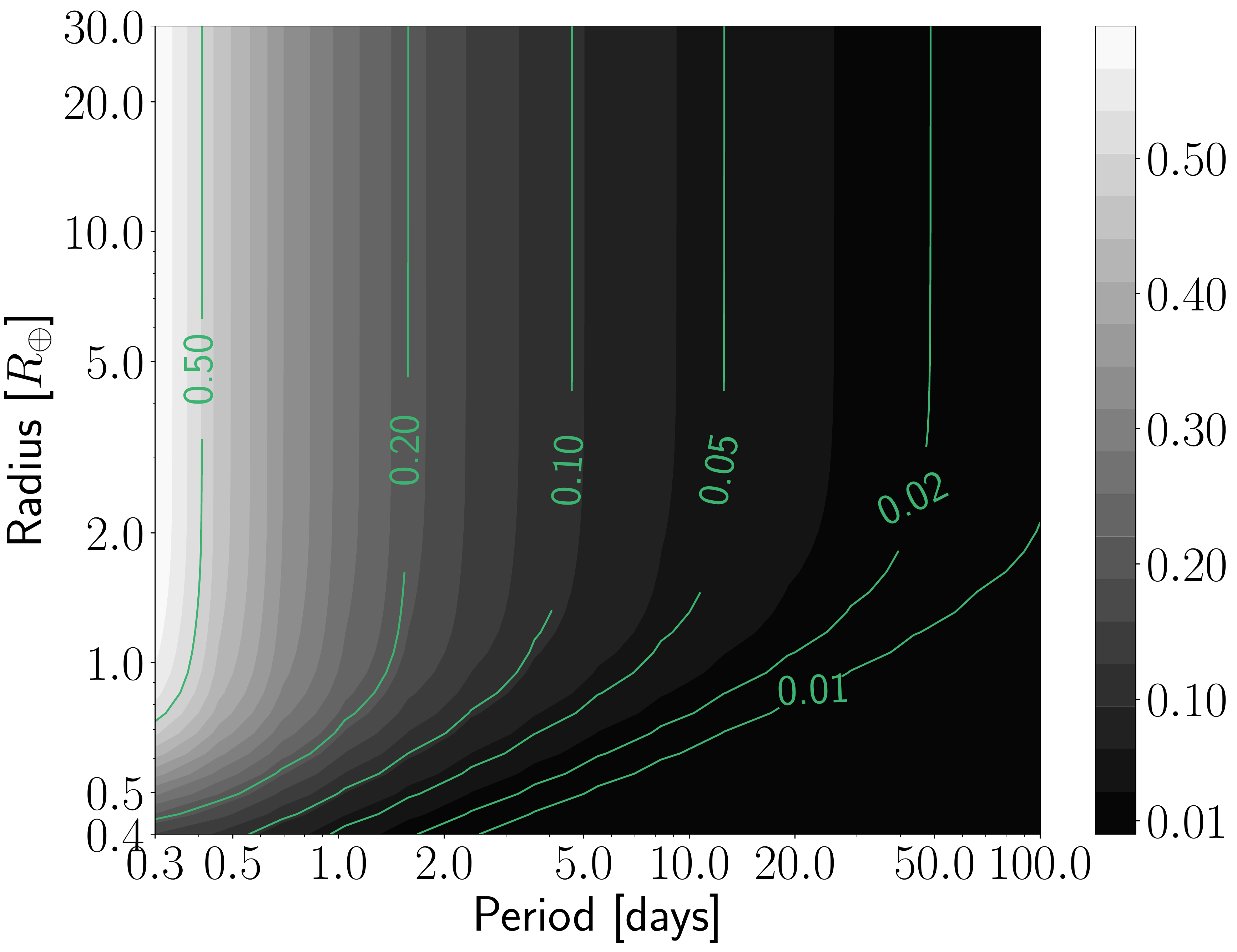}
    \caption{The average detection efficiency of the combined \textit{Kepler} Pipeline and Robovetter for our stellar sample of interest as a function of planet period and radius. The shading of the grid corresponds to the fraction of planets detected at a given radius and period. Transit probability is factored in. The colored lines show contours ranging from $1\%$ to $50\%$ of planets detected.}
    \label{detection efficiency}
\end{figure}

In measuring the detection efficiency of the \textit{Kepler} survey, we follow the procedure for \textit{Kepler} Data Release 25 as outlined in \citet{Christiansen2017Tech}, \citet{Burke&Catanzarite2017Tech}, and \citet{ThompsonEt2018ApJS}. We describe the steps briefly here, and refer the reader to the aforementioned resources for details. We first apply the same cuts to the \textit{Kepler} Q1-Q17 DR25 stellar target sample as we applied to the CKS planet candidate sample, leaving us with 30,070 stars. Following \citet{Christiansen2017Tech}, we use the pixel-level injected light curves (using only injections that match our CKS cuts) to calculate the fraction of simulated transits recovered by the pipeline as a function of the expected multiple event statistic (MES), a measure of the strength of the transit signal relative to the noise calculated by the Transiting Planet Search algorithm. We then calculate the fraction of transits correctly dispositioned as planet candidates by the Robovetter, a tool used by the \textit{Kepler} team to disposition transits as candidates or false positives, as a function of MES, as in \citet{ThompsonEt2018ApJS}. This average detection efficiency of the combined \textit{Kepler} Pipeline and the Robovetter is then fit by a gamma CDF function:

\begin{equation}
    \text{p} = F(x | a,b,c) = \frac{c}{b^a \Gamma(a)} \int^x_0{t^{a-1}e^{-t/b}dt}
\end{equation}

\noindent where $x$ is the MES, $\text{p}$ is the probability that a planet at a given MES is detected, and $a, b,$ and $c$ are coefficients that are retrieved to be $a=29.41$, $b=0.284$, $c=0.891$. We then use the KeplerPORTS\footnote{https://github.com/nasa/KeplerPORTs} Python package to calculate the detection efficiency individually for each target star in a grid of planet radius and period, using our fitted gamma CDF function to convert from estimated MES to fraction of planets detected. We then multiply this completeness of the \textit{Kepler} pipeline by the transit probability $p_\text{tra}$. Our final detection efficiency is then taken to be the average over our stellar sample, and is shown in Figure \ref{detection efficiency}.

\section{Models}\label{Models}

We constrain the planet occurrence rate density in radius, mass and period, which is defined as the number of planets per star per interval in radius, mass, and period. We denote the planet occurrence rate density by $\Gamma (P, M, R)$:

\begin{equation} \label{occ rate density (definition)}
    \Gamma (P, M, R) = \frac{dN}{dP\ dM\ dR}   \\
\end{equation}

\noindent Note that the standard set by other occurrence rate papers is to define the occurrence rate density using log intervals, but it is trivial to convert between the two by either dividing or multiplying by radius, period, and mass. We choose this definition because it simplifies the notation for the normal, log-normal and power-law parametrizations that we use herein.

In constraining the planet occurrence rate density in radius, mass and period, we must assume some parametrization. In the following sections, we present several such parametrizations that start simple and gradually add further complications. To start, we consider a model with a single population of planets. We later incorporate mixture models to account for multiple, distinct, physically motivated populations of planets.

For each parametrization, we start with the following breakdown of the occurrence rate density, $\Gamma (P, M, R)$:

\begin{equation} \label{occ rate density}
    \Gamma (P, M, R) = \Gamma_0\, \text{p}(P)\, \text{p}(M)\, \text{p}(R | M)   \\
\end{equation}

\noindent where we have separated the occurrence rate density into four components: a period distribution $p (P)$ that is independent of mass and radius, a mass distribution $p (M)$, a radius conditioned on mass distribution $p (R | M)$ given by a probabilistic mass-radius relationship, and an overall normalization term $\Gamma_0$. For all of our models, the period $P$ is assumed to be in units of days, and the planet radius $R$ and planet mass $M$ are in Earth units ($R_{\oplus}$ and $M_{\oplus}$, respectively). All period, mass and radius distributions are normalized to 1 over the boundaries we set and define later in the text.

With the use of a mixture model, Equation (\ref{occ rate density}) becomes:

\begin{equation} \label{mixture occ rate density}
    \Gamma (P, M, R) = \sum_{q=0}^{N-1} \Gamma_0\, \text{p}(P | q)\, \text{p}(M | q)\, \text{p}(R | M, q)\, \text{p}(q | M, P)   \\
\end{equation}

\noindent where $0 \leq q < N$ is the index of the mixture component with $N$ total mixture components. The mixture probability $p(q)$ may depend on mass and period, as we show in the case of envelope mass loss in section \ref{Mixture Model with Envelope Mass Loss}. In the case of $N = 1$, Equation (\ref{mixture occ rate density}) reduces to Equation (\ref{occ rate density}) above. In the following subsections, we describe four models corresponding to $N = 1, 2, 3, 4$.

For all models, we choose to parametrize the period distribution with a broken power-law, as has been done in previous studies of \textit{Kepler} occurrence rates \citep[e.g.][]{MuldersEt2018AJ} in order to account for the flattening of planet occurrence rates at around 10 days:

\begin{equation} \label{period distribution}
\begin{split}
    \text{p}(P) & =  A P^{\beta_1}, \ P < P_{\text{break}} \\
    \text{p}(P) & = A P_{\text{break}}^{\beta_1 - \beta_2} P^{\beta_2}, \ P > P_{\text{break}} \\
\end{split}
\end{equation}

\noindent where A is a normalization factor (dependent on $\beta_1$, $\beta_2$, $P_{\text{break}}$) such that the distribution is normalized to 1 between the lower and upper period bounds of 0.3 and 100 days, in line with \textit{Kepler}'s sensitivity limits. We have also modeled the period distribution as a single power-law, but found it to be a poor fit as the occurrence rate planets with short and long periods (close to the boundaries of 0.3 and 100 days, respectively) was being significantly overestimated, and the occurrence rate of planets with periods in between was consequently underestimated.

Since mass is a fundamental quantity for the structure of a planet and is one of the factors that determines a planet's radius, we choose to parametrize our joint distribution in terms of an independent mass distribution and a radius conditioned on mass distribution, instead of the converse. Despite this, the mass distribution of \textit{Kepler} planets is more uncertain than the radius or period distributions given that previous studies have relied on empirical mass-radius relations to translate \textit{Kepler} planet radii into masses \citep{MuldersEt2015bApJ, Neil&Rogers2018ApJ, Pascucci&Mulders2018ApJL}. By fitting the mass-radius relation simultaneously with the mass distribution, we aim to infer a more statistically robust mass distribution of \textit{Kepler} planets. The mass distribution has been most commonly parametrized as a broken power-law, with evidence for a break at around $\sim 20 M_{\oplus}$ for microlensing planets, which mostly orbit M dwarfs \citep{SuzukiEt2016ApJ}, and $\sim 10 M_{\oplus}$ for \textit{Kepler} planets, which mostly orbit G dwarfs \citep{Pascucci&Mulders2018ApJL, Neil&Rogers2018ApJ}. For all models, we choose to use a log-normal parametrization, which can take into account the peak-like nature of the mass distributions inferred in previous studies:

\begin{equation} \label{mass distribution}
    \text{p}(M) = \text{lnN}(M | \mu_M, \sigma_M) \\
\end{equation}

\noindent with a lower bound of $0.1 M_\oplus$, roughly corresponding to the predicted mass of the smallest \textit{Kepler} planet, and an upper bound of $10,000 M_\oplus$, well below the hydrogen burning limit. In model 4 we explore a more flexible parametrization that can reveal subtler features of the \textit{Kepler} mass distribution.

We parametrize our radius conditioned on mass distribution with a probabilistic mass-radius relationship, as first done in \citet{WolfgangEt2016ApJ}. At a given planet mass, a planet's radius is drawn from a normal distribution with a mean given by a power-law and an intrinsic scatter:

\begin{equation} \label{mr relation}
    \text{p}(R | M) = \text{N}(R | \mu (M), \sigma \mu (M)) \\
\end{equation}

\noindent where we use a fractional scatter given by $\mu \sigma$ rather than the flat scatter in \citet{WolfgangEt2016ApJ}, to account for the wide range of masses we are modeling. Following \citet{ChenKipping2017ApJ}, we separate the mass-radius relation into three regimes, corresponding to different mass ranges. Each regime has its own power-law slope as well as fractional scatter:

\begin{equation} \label{mr power-laws}
\begin{split}
    \mu_0 = \ & C M^{\gamma_0} \\
    \mu_1 = \ & C M_{\text{break,1}}^{\gamma_0 - \gamma_1} M^{\gamma_1} \\
    \mu_2 = \ & C M_{\text{break,1}}^{\gamma_0 - \gamma_1} M_{\text{break,2}}^{\gamma_1 - \gamma_2} M^{\gamma_2}
\end{split}
\end{equation}

\noindent with the subscripts {0, 1, 2} indicating the low mass range, intermediate mass range, and high mass range, respectively. Instead of abrupt transitions at the break points ${M_{\text{break,1}}, M_{\text{break,2}}}$, we use a logistic function to smooth between the different power-laws:

\begin{equation} \label{logistic functions}
\begin{split}
    S_1 = \frac{1}{1 + \exp(-5(\log M - \log M_{\text{break,1}})} \\
    S_2 = \frac{1}{1 + \exp(-5(\log M - \log M_{\text{break,2}})} \\
\end{split}
\end{equation}

\begin{equation} \label{smoothed functions}
\begin{split}
    \mu = \ & (1-S_1) \mu_0 + S_1 (1-S_2) \mu_1 + S_1 S_2 \mu_2 \\
    \sigma = \ & (1-S_1) \sigma_0 + S_1 (1-S_2) \sigma_1 + S_1 S_2 \sigma_2\\
\end{split}
\end{equation}

\noindent with $\mu$ and $\sigma$ serving as inputs to Equation (\ref{mr relation}). Finally, we introduce a lower-bound truncation to Equation (\ref{mr relation}) as in \citet{WolfgangEt2016ApJ}, taken as the radius of a pure-iron planet of a given mass. However, rather than use the analytic function fits in \citet{FortneyEt2007aApJ}, which were only fit up to $100 M_\oplus$, we use the analytic scaling relation in \citet{SeagerEt2007ApJ} (hereafter S07), which was fit up to a larger mass range appropriate for this work:

\begin{equation} \label{seager iron mass-radius relation}
\begin{split}
    R = & \ R_1 \cdot 10^{k_1 + \frac{1}{3} \log_{10}(M / M_1) - k_2 \ (M / M_1)^{k_3}} \\
    R_1 = & \ 2.52; M_1 = 5.80 \\
    k_1 = & \ {-0.20949}; k_2 = 0.0804; k_3 = 0.394 
\end{split}
\end{equation}

For a $0.1 M_{\oplus}$ planet, the corresponding radius of a pure-iron planet is $\approx 0.4 R_{\oplus}$, which we set as a hard lower bound in radius. The upper bound is taken to be $30 R_{\oplus}$, a generously inclusive upper limit based on the largest \textit{Kepler} planets.

For each model, we include a separate mass-radius relation for rocky planets, with differing implementations. Ideally, we would have a survey that is complete at the lowest planet radii of interest, and would have accurate RV masses of these small, rocky planets. While \textit{Kepler} has discovered hundreds of potentially rocky planets, the subset of these planets with RV masses remains small, and the uncertainties on the available RV masses are large (with a median uncertainty of $22\%$ in our sample). If we attempt to fit for the mass-radius relation of rocky planets (below $\sim 1.6 R_{\oplus}$ \citep{Rogers2015ApJ}), the lack of mass measurements in our sample may be problematic and lead to large uncertainties. Fortunately, the mass-radius relation of rocky planets of a given composition can be constrained from interior structure models, as in S07. Objects outside our data sample, such as Solar System planets and moons, as well as exoplanet systems with TTV mass measurements, can also constrain the rocky mass-radius relation, as done in \citet{ChenKipping2017ApJ}. The current sample of potentially rocky planets with mass and radius information exhibit a low scatter about a theoretical pure-silicate mass-radius relation \citep{DressingEt2015ApJ, ZengEt2016ApJ}. The prior information from the Solar System planets and interior structure models will be more constraining than the CKS sample used in this work. Therefore, we fix the rocky mass-radius relation to the analytic mass-radius relation of a pure silicate planet from S07:

\begin{equation} \label{seager silicate mass-radius relation}
\begin{split}
    R = & \ R_1 \cdot 10^{k_1 + \frac{1}{3} \log_{10}(M / M_1) - k_2 \ (M / M_1)^{k_3}} \\
    R_1 = & \ 3.90; M_1 = 10.55 \\
    k_1 = & \ {-0.209594}; k_2 = 0.0799; k_3 = 0.413 
\end{split}
\end{equation}

While each individual model has its own unique parametrization, with varying number of mixture populations, they all share the universal elements of the above distributions. Each model uses a broken power-law for period distributions, log-normals for mass distributions, and probabilistic broken power-laws for mass-radius relations.

\subsection{Baseline Model with a Single Population of Planets}

As our baseline model, we consider the exoplanet mass-radius-period distribution to be comprised of a single population of planets, that span a range of compositions from rocky to gas giants. This is consistent with most previous works that constrained either the mass-radius relation or radius-period distribution \citep{WolfgangEt2016ApJ, ChenKipping2017ApJ, Foreman-MackeyEt2014ApJ, Burke2015ApJ}.

For this model, our single population of planets is described by a broken power-law period distribution, a log-normal mass distribution, and a double broken power-law mass-radius relation. We replace the first equation describing the low-mass end of the mass-radius relation in Equation (\ref{mr power-laws}) with the rocky mass-radius relation from S07 in Equation (\ref{seager silicate mass-radius relation}). We fix this relation for reasons described in the preceding section, but with varying intrinsic fractional scatter $\sigma_0$, as well as the low-to-intermediate-mass transition point $M_{\text{break,1}}$. This correspondingly changes the normalization for the intermediate and high mass power-laws in Equation (\ref{mr power-laws}). We are left with a total of 13 parameters for our baseline model: $\{\Gamma_0, \beta_1, \beta_2, P_{\text{break}}, \mu_M, \sigma_M, \gamma_1, \gamma_2, \sigma_0, \sigma_1, \sigma_2, M_{\text{break,1}}, M_{\text{break,2}}\}$. In the following sections, we will refer to this model as ``model 1", since it is both the first model we present and only has one mixture component.

\subsection{Mixture Model with Envelope Mass Loss}\label{Mixture Model with Envelope Mass Loss}

A single population of planets, described by a mass-radius relation and a log-normal mass distribution, is insufficient to fully characterize the current sample of planets. Detailed analysis of the radius distribution of \textit{Kepler} planets has revealed a bimodal distribution, with a gap between $1.5-2.0 R_{\oplus}$ \citep{FultonEt2017ApJ}. This is strong evidence for two separate populations of planets (super-Earths and sub-Neptunes) that vary in envelope gas mass. This feature is not captured in the most frequently used empirical mass-radius relations \citep{WolfgangEt2016ApJ, ChenKipping2017ApJ, NingEt2018ApJ}, and thus will not be reflected in the radius or mass predictions using those relations. 

We directly model the two hypothesized populations of planets using a mixture model, building upon our baseline model introduced in the previous section. To start, we consider two populations of planets that share a common formation history and form with a primordial hydrogen/helium envelope, but one population retains their envelopes (gaseous), and the other loses their envelopes due to the incident X-ray and extreme ultraviolet flux (hereafter XUV flux) on the planet (rocky evaporated/remnant cores). These two populations are then both drawn from the same mass and period distribution, but differ by the radius conditioned on mass distribution $\text{p}(R | M, q)$. Note that what we are calling the gaseous population need not have a large H/He envelope mass fraction: a planet with $1\%$ by mass H/He has a transit radius nearly double that of a planet with the same mass but without a H/He envelope \citep[e.g.][]{Lopez&Fortney2013ApJ, Chen&Rogers2016ApJ}. In addition, for planets with massive enough envelopes, most of the H/He will not be in the gaseous phase.

For our envelope mass loss, we use the prescription for hydrodynamic, XUV-driven mass loss from \citet{LopezEt2012ApJ}. The mass-loss timescale is given by:

\begin{equation}\label{mass loss timescale}
    t_{\text{loss}} = \frac{G M^2_{\text{env}}}{\pi \epsilon R_\text{prim}^3 F_{\text{XUV,E100}}} \frac{F_{\oplus}}{F_p}
\end{equation}

\noindent where $M_{\text{env}}$ is the envelope mass of the planet, $\epsilon$ is the mass-loss efficiency, $F_{\text{XUV,E100}}$ is the XUV flux at the Earth when it was 100 My old, and $F_p$ is the bolometric incident flux on the planet. The incident flux on the planet is calculated using the planet's orbital period, as well as its host star mass, radius and temperature. The radius of the planet here, $R_\text{prim}$ is the primordial radius at formation when it had an envelope and is not necessarily the current radius of the planet. We approximate this from the mass of the planet using the median mass-radius relation of the gaseous population (i.e., the mass-radius relation in Equation (\ref{mr power-laws}) substituting in the central values of our priors for each free parameter; see Table \ref{results table} for our priors). The mass of the envelope is calculated from mass of the planet, by smoothing between an envelope mass of $M_{\rm env} = 0.1 M_p$ and $M_{\rm env}=M_p - \sqrt{M_p}$, the latter taken from the scaling of planet heavy-element mass with total mass, calculated from thermal and structural evolution models of giant planets \citep{Thorngren2016ApJ}. The smoothing is done via a logistic function as in Equation (\ref{logistic functions}), with a transition at $20 M_\oplus$, where the scaling from \citet{Thorngren2016ApJ} is valid. 

Denoting the gaseous mixture as $q=0$ and the rocky mixture as $q=1$, the probability that a planet retains its envelope is then:

\begin{equation}\label{probability of retention}
\begin{split}
   \text{p}(q=0) = & \ \text{p}_{\text{ret}} = \text{min}\left(\alpha \frac{t_{\text{loss}}}{\tau}, 1\right), \\
    \text{p}(q=1) = & \ (1 - \text{p}_\text{ret})
\end{split}
\end{equation}

\noindent where $\tau$ is the age of the star, and $\alpha$ is an additional free parameter in the model. The nominal values of $\epsilon$, $F_{\text{XUV,E100}}$ and $\tau$ we take to be $0.1$, $504 \ \text{erg} \ \text{s}^{-1} \ \text{cm}^{-2}$ and $5 \ \text{Gyr}$ for each star. In order to account for the uncertainties in these three values, we include a prefactor $\alpha$ to scale the retention probability either up or down:

\begin{equation} \label{alpha scaling}
    \alpha = \left(\frac{0.1}{\epsilon}\right) \left(\frac{504 \ \text{erg} \ \text{s}^{-1}}{F_{\text{XUV,E100}}}\right) \left(\frac{5 \ \text{Gyr}}{\tau}\right)
\end{equation}

This prescription for the envelope retention may seem fine-tuned, given the numerous assumptions we made. Our goal here is just to include a reasonable, simple model of mass loss, understanding that there may be complexity that we are not accounting for. We allow $\alpha$ to vary to partially account for this uncertainty. The photoevaporation that we model here has been shown to be able to reproduce the Fulton radius gap by itself \citep{VanEylenEt2018MNRAS,Fulton&Petigura18AJ}. However, there are other mass loss mechanisms that can reproduce the radius gap without the need for photoevaporation, such as core-powered mass loss \citep{GinzburgEt2017MNRAS, Gupta&Schlichting2018Arxiv}. We do not include multiple mass loss mechanisms as it would unnecessarily complicate the model and leave exploring additional mass loss prescriptions to future work.

As previously mentioned, our two populations both formed with a gaseous envelope, but one population lost their envelopes through hydrodynamic mass loss and are currently rocky, and the other population retained their envelopes and are currently gaseous. For this model and subsequent models, we introduce the following subscripts: ``fr" to represent formed rocky, ``fg" to represent formed gaseous, ``cr" to represent currently rocky, and ``cg" to represent currently gaseous. Our two populations differ in the radius conditioned on mass distribution $\text{p}(R | M, q)$ only. Thus, using Equation (\ref{mixture occ rate density}), our final mixture model rate density becomes:

\begin{equation} \label{mass loss final rate density}
\begin{split}
    \text{p}(R | M, q=0) = & \ \text{p}(R | M)_\text{cg} \\
    \text{p}(R | M, q=1) = & \ \text{p}(R | M)_\text{cr} \\
    \Gamma (P, M, R) = & \ \Gamma_0\, \text{p}(P)_\text{fg}\, \text{p}(M)_\text{fg}\, \big[\text{p}_{\text{ret}} \ \text{p}(R | M)_\text{cg} \\
                     & \ +  (1 - \text{p}_{\text{ret}}) \text{p}(R | M)_\text{cr} \big] \\
\end{split}
\end{equation}

We use Equation (\ref{seager silicate mass-radius relation}) from S07 for the rocky component of the mass-radius relation $\text{p}(R | M)_\text{cr}$, with a fixed $5\%$ scatter. The conditional distribution then becomes:

\begin{equation} \label{rocky mixture r(m) relation}
    \text{p}(R | M)_\text{cr} = N(R_{\text{S07}}(M), 0.05\cdot R_{\text{S07}}(M))
\end{equation}

For the gaseous mixture component mass-radius relation $\text{p}(R | M)_\text{cg}$, we use Equation (\ref{mr power-laws}), allowing $C$ and $\gamma_0$ to vary along with the scatter $\sigma_0$. With the low amount of mass-radius information in the dataset at these low masses, we don't expect the model to tightly constrain these parameters, but we value flexibility by including them. In total, our mixture model has 16 free parameters: $\{\Gamma_0, \beta_1, \beta_2, P_{\text{break}}, \mu_M, \sigma_M, C, \gamma_0, \gamma_1, \gamma_2, \sigma_0, \sigma_1, \sigma_2,$ \newline $M_{\text{break,1}}, M_{\text{break,2}}, \alpha\}$. We will refer to this model with two mixture components as ``model 2" in the following sections.

\subsection{Two Populations of Rocky Planets}\label{Two Populations of Rocky Planets}

Our previous model assumed a shared formation history for the two populations of planets. Next, we allow for the possibility that the rocky and gaseous planets arise from different formation scenarios. This manifests as the sub populations having distinct mass and period distributions. We still maintain the envelope mass loss introduced in the previous model, leaving us with three populations of planets: an ``intrinsically rocky" population, a population of evaporated cores (which lost their envelopes through hydrodynamic mass loss), and a population of planets with H/He envelopes. The intrinsically rocky and evaporated cores share a common radius conditioned on mass distribution, and the evaporated cores and gaseous envelope planets arise from a shared underlying mass and period distribution (though due to the dependence of envelope mass loss on mass and period, their mass and period distributions will be different). Following the notation from the preceding section, we denote planets belonging to the ``intrinsically rocky" population as $q=2$. Instead of modeling the mixture probability $\text{p}(q=2)$ using physics, we can simply model the prior $p(q)$ as a 2-simplex:

\begin{equation} \label{mixture q prior}
\begin{split}
    \text{p}(0,1) = \ & 1 - Q_\text{fr} \\
    \text{p}(2) = \ & Q_\text{fr}
\end{split}
\end{equation}

\noindent where $0 < Q_\text{fr} < 1$ is an additional parameter in our model parameterizing the total fraction of intrinsically rocky planets compared to ``intrinsically gaseous" planets ($q=0,1$). Note that $\text{p}(q=0)$ and $\text{p}(q=1)$ are further modified by the envelope retention from the previous model, becoming:

\begin{equation} \label{model 3+ mixture q=0,1 prior}
\begin{split}
    \text{p}(0) = \ & (1 - Q_\text{fr}) \ \text{p}_\text{ret}\\
    \text{p}(1) = \ & (1 - Q_\text{fr}) \ (1 - \text{p}_\text{ret})
\end{split}
\end{equation}

Following a similar expansion as with the previous model, we obtain the final rate density:

\begin{equation} \label{mixture final rate density}
\begin{split}
    \Gamma (P, M, R) = & \ \Gamma_0\, \Big[ (1-Q_\text{fr})\ \text{p}(P)_\text{fg}\, \text{p}(M)_\text{fg}\, \big[ \text{p}_\text{ret}\ \text{p}(R | M)_\text{cg} \Big. \big.\\
                     & \ \big. + (1-\text{p}_\text{ret}) \text{p}(R | M)_\text{cr} \big] \\
                     & \ \Big. + \ Q_\text{fr} \ \text{p}(P)_\text{fr}\, \text{p}(M)_\text{fr}\, \text{p}(R | M)_\text{cr} \vphantom{\text{p}_\text{ret}}\Big]
\end{split}
\end{equation}

\noindent For our intrinsically rocky population, we use the same parametrization of the period and mass distributions:

\begin{equation} \label{rocky mixture period/mass distributions}
\begin{split}
    \text{p}(P)_\text{fr} = & A_\text{fr} P^{\beta_{1,\text{fr}}}, \ P < P_{\text{break,fr}} \\
    \text{p}(P)_\text{fr} = & A_\text{fr} P_{\text{break,fr}}^{\beta_{1,\text{fr}} - \beta_{2,\text{fr}}} P^{\beta_{2,\text{fr}}}, \ P > P_{\text{break,fr}} \\
    \text{p}(M)_{\text{fr}} = & \ \text{lnN}(M | \mu_{M,\text{fr}}, \sigma_{M,\text{fr}}) \\
\end{split}
\end{equation}

\noindent with $A_\text{fr}$ as a normalization constant dependent on $\beta_{1,\text{fr}}, \beta_{2,\text{fr}}$ and $P_{\text{break,fr}}$ as in Equation (\ref{period distribution}). 

After adding the the three parameters from the period distribution, the two parameters from the mass distribution, and the fraction of intrinsically rocky planets $Q_\text{fr}$, this model has a total of 22 free parameters. Following suit with our previous models, we will refer to this model with three populations of planets as ``model 3".

\subsection{Mixture of Log-Normals}\label{Mixture of Log-Normals}

Perhaps the biggest weakness of the previous models is the lack of flexibility in the mass distribution. By assuming a log-normal distribution, we are potentially missing features in the \textit{Kepler} mass distribution that can't be captured with a two-parameter distribution. 

We modify our mixture model as described in the previous section by using a mixture of log-normals as the mass distribution of the gaseous population. In practice, this is implemented by extending the previous model to include additional mixtures, modeling $\text{p}(q)$ as an ($N-1$)-simplex:

\begin{equation} \label{N mixture q prior}
\begin{split}
    \text{p}(q) = \ & Q_q \\
    \sum_{q=0}^{N-2} Q_q = \ & 1
\end{split}
\end{equation}

\noindent where $q$ is the $q$th mixture and $Q_q$ is the fraction of planets belonging to that mixture. While technically this mixture of log-normals is implemented as additional mixtures in the model, qualitatively the multiple mixtures are still considered to be part of intrinsically gaseous population, so we still consider this to be a model with three populations of planets. In this paper we try $N=4$, which corresponds to a intrinsically gaseous population of planets drawn from a mixture of two log-normals, but higher $N$ can be easily modeled. Models 1, 2, and 3 described in the preceding sections correspond to $N=1,2,3$ respectively. Our model with $N=4$ requires three additional parameters: the mean and spread of the second component of the intrinsically gaseous mass distribution, $\mu_{M,2}$ and $\sigma_{M,2}$, as well as the overall fraction of planets belonging to this component, $Q_2$. This model has a total of 25 parameters and is the final model we present in this paper, labeled ``model 4".

\section{Fitting}\label{Fitting}

\setlength\extrarowheight{3pt}
\begin{table*}
    \centering
    \begin{tabular}{c c c c c c c}
    \hline
    Parameter & Units & Model 1 & Model 2 & Model 3 & Model 4 & Prior \\
    \hline\hline
    $\Gamma_0$ & $N_\text{pl} / N_\text{s}$ & $1.28 \substack{+0.06 \\ -0.06}$ & $1.14 \substack{+0.06 \\ -0.05}$ & $1.04 \substack{+0.05 \\ -0.05}$ & $1.05 \substack{+0.05 \\ -0.05}$ & $\ln$\text{N}(0, 1) \\
    $C$ & $R_{\oplus}$ & - & $2.37 \substack{+0.20 \\ -0.22}$ & $2.08 \substack{+0.28 \\ -0.23}$ & $2.32 \substack{+0.31 \\ -0.34}$ & \text{N}(2.5, 1) \\
    $\gamma_0$ & - & - & $0.00 \substack{+0.05 \\ -0.05}$ & $0.08 \substack{+0.06 \\ -0.07}$ & $0.02 \substack{+0.09 \\ -0.07}$ & \text{N}(0, 0.1) \\
    $\gamma_1$ & - & $0.42 \substack{+0.01 \\ -0.01}$ & $0.74 \substack{+0.05 \\ -0.04}$ & $0.75 \substack{+0.05 \\ -0.05}$ & $0.61 \substack{+0.04 \\ -0.03}$ & \text{N}(0.6, 0.1) \\
    $\gamma_2$ & - & $0.08 \substack{+0.07 \\ -0.07}$ & $0.04 \substack{+0.06 \\ -0.06}$ & $0.04 \substack{+0.05 \\ -0.06}$ & $-0.01 \substack{+0.05 \\ -0.05}$ & \text{N}(0, 0.1) \\
    $\sigma_0$ & - & $0.07 \substack{+0.02 \\ -0.01}$ & $0.18 \substack{+0.02 \\ -0.02}$ & $0.16 \substack{+0.02 \\ -0.02}$ & $0.16 \substack{+0.02 \\ -0.02}$ & $\ln$\text{N}(-1.8, 0.25)* \\
    $\sigma_1$ & - & $0.27 \substack{+0.03 \\ -0.03}$ & $0.34 \substack{+0.06 \\ -0.05}$ & $0.33 \substack{+0.06 \\ -0.05}$ & $0.23 \substack{+0.05 \\ -0.04}$ & $\ln$\text{N}(-1.3, 0.25) \\
    $\sigma_2$ & - & $0.11 \substack{+0.03 \\ -0.02}$ & $0.10 \substack{+0.02 \\ -0.02}$ & $0.10 \substack{+0.02 \\ -0.02}$ & $0.11 \substack{+0.02 \\ -0.02}$ & $\ln$\text{N}(-2.3, 0.25) \\
    $M_{\text{break},1}$ & $M_{\oplus}$ & $0.43 \substack{+0.08 \\ -0.08}$ & $17.4 \substack{+2.5 \\ -2.3}$ & $20.0 \substack{+3.4 \\ -3.0}$ & $9.76 \substack{+1.94 \\ -1.43}$ & $\ln$\text{N}(2, 1) \\
    $M_{\text{break},2}$ & $M_{\oplus}$ & $267.08 \substack{+51.6 \\ -38.7}$ & $175.7 \substack{+25.3 \\ -20.5}$ & $168.9 \substack{+24.0 \\ -19.0}$ & $162.2 \substack{+24.9 \\ -20.1}$ & $\ln$\text{N}(5, 0.25) \\
    $\mu_M$ & $\ln(\frac{M}{M_\oplus})$ & $0.29 \substack{+0.14 \\ -0.16}$ & $1.00 \substack{+0.07 \\ -0.08}$ & $1.72 \substack{+0.13 \\ -0.14}$ & $0.60 \substack{+0.49 \\ -0.47}$ & \text{N}(1, 2)* \\
    $\sigma_M$ & $\ln(\frac{M}{M_\oplus})$ & $1.72 \substack{+0.09 \\ -0.07}$ & $1.65 \substack{+0.06 \\ -0.06}$ & $1.38 \substack{+0.08 \\ -0.07}$ & $2.39 \substack{+0.29 \\ -0.24}$ & $\ln$\text{N}(1, 0.25)* \\
    $\mu_{M,2}$ & $\ln(\frac{M}{M_\oplus})$ & - & - & - & $1.72 \substack{+0.08 \\ -0.08}$ & \text{N}(2, 0.5) \\
    $\sigma_{M,2}$ & $\ln(\frac{M}{M_\oplus})$ & - & - & - & $0.63 \substack{+0.07 \\ -0.06}$ & $\ln$\text{N}(0, 0.25) \\
    $\mu_{M,\text{fr}}$ & $\ln(\frac{M}{M_\oplus})$ & - & - & $-0.15 \substack{+0.25 \\ -0.34}$ & $-0.31 \substack{+0.24 \\ -0.31}$ & \text{N}(0, 2) \\
    $\sigma_{M,\text{fr}}$ & $\ln(\frac{M}{M_\oplus})$ & - & - & $1.61 \substack{+0.18 \\ -0.14}$ & $1.42 \substack{+0.19 \\ -0.15}$ & $\ln$\text{N}(0.5, 0.25) \\
    $\beta_1$ & - & $0.90 \substack{+0.08 \\ -0.07}$ & $0.88 \substack{+0.07 \\ -0.07}$ & $1.24 \substack{+0.19 \\ -0.17}$ & $1.14 \substack{+0.15 \\ -0.15}$ & \text{N}(0.5, 0.5) \\
    $\beta_2$ & - & $-0.69 \substack{+0.06 \\ -0.05}$ & $-0.76 \substack{+0.06 \\ -0.06}$ & $-0.71 \substack{+0.07 \\ -0.06}$ & $-0.73 \substack{+0.06 \\ -0.07}$ & \text{N}(-0.5, 0.5) \\
    $P_{\text{break}}$ & \text{days} & $7.01 \substack{+0.44 \\ -0.49}$ & $7.16 \substack{+0.45 \\ -0.44}$ & $8.09 \substack{+0.71 \\ -0.57}$ & $7.81 \substack{+0.59 \\ -0.58}$ & $\ln$\text{N}(2, 1) \\
    $\beta_{1,\text{fr}}$ & - & - & - & $0.90 \substack{+0.17 \\ -0.15}$ & $0.82 \substack{+0.17 \\ -0.18}$ & \text{N}(0.5, 0.5) \\
    $\beta_{2,\text{fr}}$ & - & - & - & $-1.25 \substack{+0.19 \\ -0.26}$ & $-1.11 \substack{+0.19 \\ -0.24}$ & \text{N}(-0.5, 0.5) \\
    $P_{\text{break},\text{fr}}$ & \text{days} & - & - & $4.38 \substack{+0.60 \\ -0.44}$ & $4.58 \substack{+1.06 \\ -0.60}$ & $\ln$\text{N}(2, 1) \\
    $\alpha$ & - & - & $7.98 \substack{+1.40 \\ -1.36}$ & $7.29 \substack{+1.84 \\ -1.52}$ & $8.39 \substack{+1.53 \\ -1.45}$ & $\ln$\text{N}(0, 1) \\
    $Q_{\text{fr}}$ & - & - & - & $0.20 \substack{+0.06 \\ -0.05}$ & $0.21 \substack{+0.05 \\ -0.05}$ & \text{U}(0, 1)* \\
    $Q_2$ & - & - & - & - & $0.26 \substack{+0.07 \\ -0.06}$ & \text{D}(1) \\
    \hline
    \end{tabular}
    \caption{The retrieved median and 1-$\sigma$ intervals for each parameter of interest for each model, as well the units and the assumed prior. N represents a normal distribution, $\ln$N represents a lognormal distribution, U represents a uniform distribution, and D represents a Dirichlet distribution with parameter length equal to the number of mixture components. Descriptions of each parameter can be found in section 3. Parameters with units listed as `-' are dimensionless. The priors with an * differ depending on the model. In model 1, the $\sigma_0$ parameter takes a $\ln\text{N}(-2.8, 0.25)$ prior to emulate the low scatter in the mass-radius relation for small, rocky planets. The $\mu_M$ and $\sigma_M$ priors are changed to $\text{N}(0, 0.5)$ and $\ln \text{N}(0, 0.25)$ respectively in model 4 in order to impose identifiability in the mixture model by clearly separating the two mass distribution components. In addition for model 4, the parameter $Q_\text{fr}$ is fit together with the other mixture components as a simplex and is assigned a Dirichlet prior that gives any combination of mixture probabilities equal weighting.}
    \label{results table}
\end{table*}

We model our \textit{Kepler} planet catalog as draws from an inhomogeneous Poisson process. This technique has been used previously to constrain the occurrence rate of \textit{Kepler} planets in radius-period space \citep{Foreman-MackeyEt2014ApJ}. We use a corrected form of their likelihood:

\begin{equation} \label{inhom. poisson process likelihood}
    \text{p}(\{w_k\} | \theta) = \text{exp} \left(- \int \hat{\Gamma}_{\theta} (w) \text{d}w \right) \prod_{k=1}^K \Gamma_\theta (w_k) \text{p}_\text{tr} (w_k)
\end{equation}

\noindent where $w$ represents the planet parameters of interest (in our case the true planet mass, $M_\text{true}$, the true planet radius $R_\text{true}$, and the planet period $P$), $k$ is the index for a particular planet for a total of $K$ planets, $\left\{w_k\right\}$ is the collection of all planet properties, $\theta$ represents our population-level parameters, $p_\text{tr}$ is the a priori probability of a planet to transit ($R_\text{star}/a$) and $\hat{\Gamma}_{\theta}$ is the observable occurrence rate density:

\begin{equation} \label{observable occurrence rate density}
    \hat{\Gamma}_{\theta} (w) = \Gamma_\theta (w) \text{p}_\text{det} (w) \text{p}_\text{tr} (w) 
\end{equation}

\noindent with $p_\text{det}$ representing the probability that a planet of a given radius and period would be detected by the \textit{Kepler} pipeline and subsequently identified as a planet candidate by the Robovetter. The true planet mass and radius, $M_\text{true}$ and $R_\text{true}$ respectively, are considered to be free parameters in the model. As in \citet{Foreman-MackeyEt2014ApJ}, we expand the likelihood to account for the fact that our inputs are observed masses and radii, rather than the true values:

\begin{equation} \label{expanded inhom. poisson process likelihood}
\begin{split}
    \text{p}(\{x_k\} | \theta, \{w_k\}) = & \ \text{exp} \left(- \int \hat{\Gamma}_{\theta} (w) \text{d}w \right) \\
    & \ \cdot \prod_{k=1}^K \Gamma_\theta (w_k) \text{p}(x_k | w_k) \text{p}_\text{tr} (w_k)
\end{split}
\end{equation}

\noindent where $x_k$ represents the measurements of the $k$th planet's mass and radius. The $\text{p}(x_k | w_k)$ term is given by the following:

\begin{equation} \label{true mass and radius}
\begin{split}
    \text{p}(x_k | w_k) = & \ N(R_{\text{true},k} | R_{\text{obs},k}, \sigma_{R,\text{obs},k}) \\
     & \ \cdot N(M_{\text{true},k} | M_{\text{obs},k}, \sigma_{M,\text{obs},k})
\end{split}
\end{equation}

\noindent where $R_\text{obs,k}, \sigma_\text{R,obs,k}$ are the measured radius and radius uncertainty for the $k$th planet, and similarly for the measured masses. In the case where a planet does not have a mass measurement, Equation (\ref{true mass and radius}) changes to only include the first term. Note that due to the high precision to which periods of \textit{Kepler} planets are measured, we neglect observational uncertainty in the orbital period and take $P_\text{obs,k} = P_\text{true,k}$

The difference between the likelihood used here and that presented in \citet{Foreman-MackeyEt2014ApJ} is the absence of the $p_\text{det}$ term in the multiplicative sum. Including the $p_\text{det}$ term in both the integral and the individual planet likelihoods is incorrectly conditioning on the detection twice \citep{Loredo04AIPCS, MandelEt19MNRAS}. The geometric  transit probability, $p_\text{tr}$, appears in both because the subset of planets that are transiting is the population we are able to detect.

For our mixture models, we use a modified form of the inhomogeneous Poisson process likelihood:

\begin{equation} \label{mixture likelihood full}
\begin{split}
    \text{p}(\{w_k\} | \theta) = & \ \text{exp}  \left(- \int \left( \sum_{q=0}^{N-1} \text{p}(q) \hat{\Gamma}_{\theta,q} (w) \right) \text{d}w \right) \\
    & \ \cdot \prod_{k=1}^K \left( \sum_{q=0}^{N-1} \text{p}(q) \Gamma_{\theta,q} (w_k) \text{p}_\text{tr} (w_k) \right) \\
\end{split}
\end{equation}

\noindent where $\text{p}(q)$ are the mixture probabilities of the $q$th mixture, for a total of $N$ mixtures.

Given that a minority of transiting planets in our sample have RV mass measurements, and given the complexity of some of the models we are attempting to fit, we are forced to adopt reasonably informative priors for our population level model parameters. We refer to the literature to determine our priors where previous work has been done (without explicitly using past results as our priors), striking a balance between flexibility and being informative. Our priors for each parameter in our models are listed in the last column of Table \ref{results table}. 

To fit our models to the data, we use the Python implementation of the Stan statistical software package \citep{CarpenterEt2017}\footnote{http://mc-stan.org}. Stan uses the No-U-Turn Sampler (NUTS) MCMC algorithm, a method of numerically evaluating hierarchical Bayesian models. The ability of NUTS to efficiently handle large dimensional spaces makes it particularly suitable for this work. The CKS sample contains over a thousand planets, and each is modeled with a ``true" mass and radius, totalling over two thousand parameters.  For each model, we ran 8 chains each with 3,000 iterations. The first 2,000 iterations of each chain are used for warm up, where they are allowed to reach their equilibrium distributions, and the MCMC algorithm fine-tunes its internal parameters. We are left with 8,000 posterior samples of each parameter. To assess the convergence and independence of each chain, we look at the Gelman-Rubin convergence diagnostic, $\hat{R}$. For each parameter, we ensure that $\hat{R} < 1.01$, a reasonable benchmark for chains mixing well.

In order to assess the validity of our model fits, we perform tests on simulated data for each of our models. For each model, we generate 10 simulated catalogs per set of chosen model parameters. We ensure that the model retrieval for each parameter is centered on the true value, without being biased high or low. We repeat this for several sets of model parameters drawn from our prior distributions to ensure we can sufficiently distinguish between these different parameters using the same priors. 

Since the inhomogeneous Poisson process likelihood in Eqs (\ref{inhom. poisson process likelihood}) and (\ref{mixture likelihood full}) contains a three-dimensional integral, the computational speed is severely bottlenecked by the grid of period, mass and radius points we use to compute the integral, especially in the mixture model case where the mass and radius distributions are no longer independent of period. We use a grid of size 11x41x41 in period-mass-radius space, with a detection efficiency grid of the same size to compute this integral. This grid size was chosen based on the condition of minimizing both the run time and error; we keep errors down to a few percent when compared to using a denser grid.

\section{Results}\label{Results}

\subsection{Model Fits}\label{Model Fits}

\begin{figure}
    \centering
    \includegraphics[width=1\columnwidth]{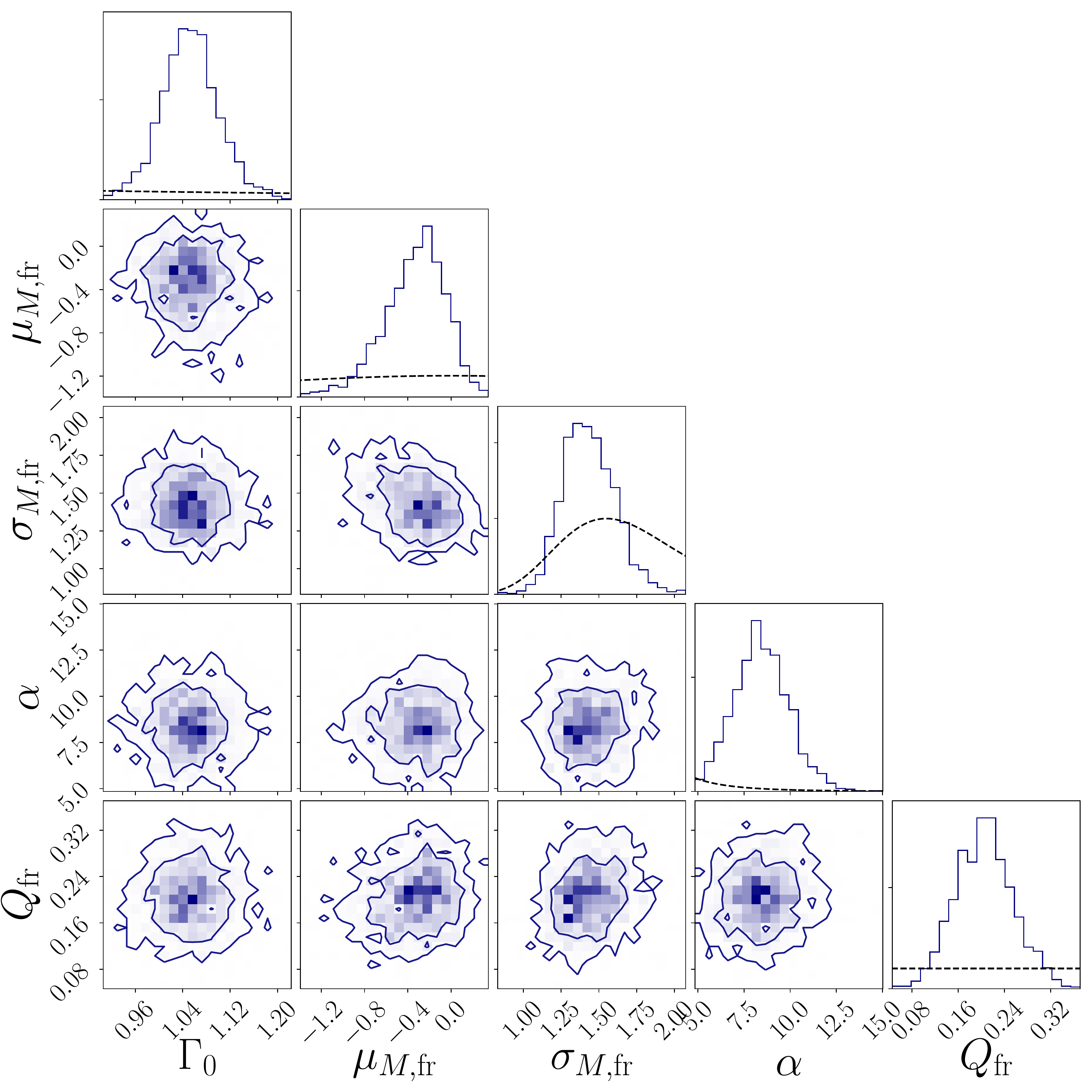}
    \caption{The 2D posteriors for a selection of parameters constraining the rocky mixture with model 4. The parameters are: $\Gamma_0$, the overall occurrence rate normalization (planets per star); $\mu_{M,\text{fr}}$, the mean of the log-normal distribution for the intrinsically rocky population; $\sigma_{M,\text{fr}}$, the spread of the log-normal distribution for the intrinsically rocky population; $\alpha$, the scaling of the mass-loss equation; $Q_\text{fr}$, the total fraction of intrinsically rocky planets. The contours are at the $1\sigma$ and $2\sigma$ levels, corresponding to the $68\%$ and $95\%$ confidence intervals, with the shading indicating the density of posterior samples in a given bin. The black dotted lines in the 1D distributions show the assumed priors.}
    \label{rocky mixture cornerplot}
\end{figure}

Our results are summarized in Table \ref{results table}, where we present the median and $1\sigma$ uncertainties ($68\%$ confidence interval) for each parameter, for each of our 4 models, along with the assumed prior. Of course, to understand the full picture, one must look at the full N-dimensional posterior to account for correlations between parameters rather than looking at each parameter independently. For all of the following results and plots, we sample from the posteriors of each of our models such that the uncertainties from the model fits are folded in.

For model 1, the best comparison for the mass-radius portion of the model is to \citet{ChenKipping2017ApJ} (hereafter CK17), as we have modeled our mass-radius relation after their work. We find a shallower mid-size slope $\gamma_1$ ($0.42 \substack{+0.01 \\ -0.01}$ in our work vs $0.59 \substack{+0.04 \\ -0.03}$ in CK17) with a slightly positive slope for $\gamma_2$ ($0.08 \substack{+0.07 \\ -0.07}$ vs $-0.04 \substack{+0.02 \\ -0.02}$), in the giant planet regime. Our breaks in the mass-radius relation are found at significantly lower ($0.43 \substack{+0.08 \\ -0.08}$ vs $2.04 \substack{+0.66 \\ -0.59}\ M_\oplus$ for $M_{\text{break,1}}$) and higher ($267 \substack{+52 \\ -39}$ vs $132 \substack{+18 \\ -21}\ M_\oplus$ for $M_{\text{break,2}}$) masses. This low value for the low mass break is below most planets in the sample, perhaps indicating weak support for including this break based on the CKS sample. This is supported by the nonparametric mass-radius relation fit by \citet{NingEt2018ApJ}, which found no evidence for a low mass break in the mass-radius relation with the current mass-radius sample. The scatter parameters $\sigma_0$, $\sigma_1$, $\sigma_2$ that we retrieve are higher across the board ($0.07 \substack{+0.02 \\ -0.01}$ vs $0.04 \substack{+0.01 \\ -0.01}$, $0.27 \substack{+0.03 \\ -0.03}$ vs $0.15 \substack{+0.02 \\ -0.01}$, and $0.11 \substack{+0.03 \\ -0.02}$ vs $0.07 \substack{+0.01 \\ -0.01}$, respectively). For each of our parameters, the uncertainty is much higher than in CK17, likely due to the much more limited sample of planets with mass and radius measurements that we condition on. Whereas we only include masses for planets in the CKS sample with RV mass measurements (described in Section 2), CK17 includes many more giant planets and solar system planets, among others. The fact that we are modeling the full 3D distribution and are accounting for non-detections and selection effects while CK17 simply modeled the mass conditioned on radius distribution also likely widens the posteriors for these parameters. We note that the goal of this work to fit the mass-radius relation together with the mass and period distributions self-consistently using one dataset, and the mass-radius relation is just one aspect of this overall distribution.

The overall normalization (planets per star) within our range of parameter space ($0.4 < \frac{R}{R_{\oplus}} < 30$ , $0.3 < \frac{P}{\text{1 day}} < 100, 0.1 < \frac{M}{M_{\oplus}} < 10,000$), $\Gamma_0$, is found to be $1.28 \substack{+0.06 \\ -0.06}$ for model 1 (for continued discussion of occurrence rates, see section 5.2). The log-normal mass distribution peaks at a value of $\mu_M = 0.29 \substack{+0.14 \\ -0.16}$ (corresponding to $1.34 M_{\oplus}$) with a spread in natural log space of $\sigma_M = 1.72 \substack{+0.09 \\ -0.07}$. Finally, the period distribution has a break at $7.0 \substack{+0.44 \\ -0.49}$ days with a low period power-law slope of $0.90 \substack{+0.08 \\ -0.07}$ and a high period power-law slope of $-0.69 \substack{+0.06 \\ -0.05}$. \citet{MuldersEt2018AJ}, who also parametrized the period distribution of \textit{Kepler} planets with a broken power-law, found a break at $12 \substack{+3 \\ -2}$ days and slopes of $1.5 \substack{+0.5 \\ -0.3}$ and $0.3 \substack{+0.1 \\ -0.2}$. However, they defined their occurrence rate density in terms of $\log P \log R$, which means their slopes correspond to slopes of $0.5$ and $-0.7$ in our model. Our slopes are thus compatible within uncertainties, noting that they consider a period range of 2-400 days, compared to our range of 0.3-100 days, and parametrize the rest of the model differently, which could account for any differences.

Introducing XUV-driven hydrodynamic mass loss with model 2 introduces several new parameters, and also significantly affects several existing model parameters. Replacing the fixed low-mass mass-radius relation in model 1, we fit for the slope $\gamma_0$ and normalization $C$ of this low-mass gaseous planet regime. We find a nearly flat slope of $0.04 \substack{+0.06 \\ -0.06}$, a normalization of $2.37 \substack{+0.20 \\ -0.22}\ R_\oplus$, a first break in the mass-radius relation at $17.4 \substack{+2.5 \\ -2.3}\ M_\oplus$, and a scatter of $0.18 \substack{+0.02 \\ -0.02}$. Given the flat slope, this normalization should correspond to the $2.4\ R_\oplus$ peak of the bimodal radius distribution first uncovered in \citet{FultonEt2017ApJ}; indeed, we find our value to be consistent with their reported location of the second peak. Allowing the possibility of envelope mass loss shifts the intermediate-mass slope $\gamma_1$ of the mass-radius relation to $0.74 \substack{+0.05 \\ -0.04}$, the peak of the log-normal mass distribution to higher mass ($1.00 \substack{+0.07 \\ -0.08}$, or $2.72\ M_\oplus$), and the second break in the mass-radius relation to lower mass ($175.7 \substack{+25.3 \\ -20.5}\ M_\oplus$). Finally, we allowed the prefactor governing envelope mass loss to vary, and found that it favored easier envelope retention than our assumed values for stellar age, mass loss efficiency, and XUV flux from the star ($\alpha = 7.98 \substack{+1.40 \\ -1.36}$).

With model 3 we add a population of intrinsically rocky planets with separate mass and period distributions from the gaseous and evaporated core populations in model 2. Compared to the gaseous and evaporated core populations, the intrinsically rocky population has a log-normal mass distribution that peaks at a lower mass ($\mu_{M,\text{fr}} = -0.15 \substack{+0.25 \\ -0.34}$ vs $\mu_M = 1.72 \substack{+0.13 \\ -0.14}$, or $0.86\ M_\oplus$ vs $5.58\ M_\oplus$), and a period distribution with a shallower slope at low periods ($\beta_{1,\text{fr}} = 0.90 \substack{+0.17 \\ -0.15}$ vs $\beta_1 = 1.24 \substack{+0.19 \\ -0.17}$) and steeper slope at high periods ($\beta_{2,\text{fr}} = -1.25 \substack{+0.19 \\ -0.26}$ vs $\beta_2 = -0.71 \substack{+0.07 \\ -0.06}$), leading to more planets at shorter periods. While the overall period distribution for the intrinsically rocky population is shifted towards shorter periods, there are also more intrinsically rocky planets at longer orbital periods compared to the evaporated core population, as most of these longer period intrinsically gaseous planets retain their envelopes instead of becoming evaporated cores. Adding this additional population does not significantly affect the mass-radius relation for the gaseous population nor the $\alpha$ factor for the envelope mass loss. We find the total fraction of intrinsically rocky planets to be $Q_\text{fr} = 0.20 \substack{+0.06 \\ -0.05}$. While this intrinsically rocky population is in many ways degenerate with the evaporated core population, the fact that it is well constrained above zero perhaps indicates the need for rocky planets either at higher masses or longer periods than the mass loss prescription would allow. For further discussion of the need for the intrinsically rocky population and the meaning of these labels, see Section \ref{Mixture Labels}. We include a subset of the 2D posteriors for model 3 in Figure \ref{rocky mixture cornerplot}, for the parameters relevant to the intrinsically rocky distribution. We find that the overall normalization is not strongly correlated with any of these parameters. However, there exists a weaker correlation with the mass loss scaling factor $\alpha$: higher values, which correspond to higher envelope retention probabilities, lead to a higher fraction of intrinsically rocky planets $Q_\text{fr}$, to balance with the fewer number of evaporated cores. This is evidence for the degeneracy between the evaporated core and intrinsically rocky populations

\begin{figure*}[ht]
    \centering
    \includegraphics[width=0.8\textwidth]{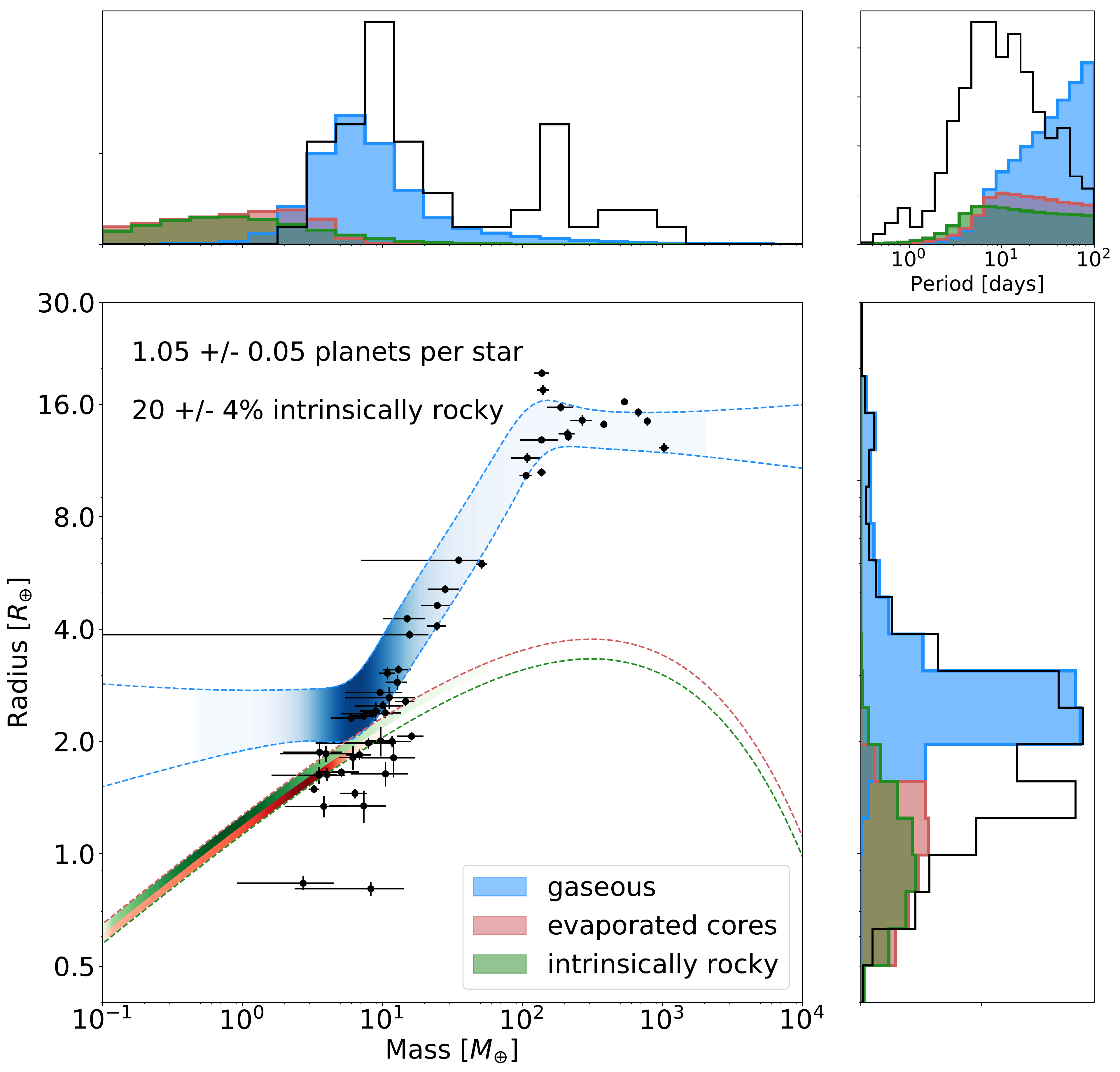}
    \caption{The joint Mass-Radius-Period distribution using our model 4 which has three mixture populations of planets. Our results for the gaseous population are shown in blue, evaporated cores in red, and intrinsically rocky in green. In the bottom left panel, we show the mass-radius relation. The dotted lines indicate the central $68\%$ of planets drawn at a given mass, and the shading indicates the occurrence at that mass relative to the overall occurrence of that mixture. The evaporated core and intrinsically rocky mixtures are stacked on top of each other in the mass-radius plane for clarity; both follow the same mass-radius relation and the central $68\%$ of planets in these two mixtures fall between the red and green dotted lines. The \textit{Kepler} planets in our sample with RV mass measurements are shown in black with their respective measurement errors. At the top we show the underlying mass distributions of the three populations, as well as the distribution of mass measurements in our sample in black. Similarly, we show the underlying radius distributions on the right with the radius distribution of the \textit{Kepler} sample in black, and we show the period distributions in the top right.}
    \label{Model 4 Main Plot}
\end{figure*}

In model 4, we allow a more flexible mass distribution for the gaseous and evaporated core populations by using a mixture of two log-normal distributions. We find that the mean of the high mean mass log-normal in this model is close to that of model 3 ($\mu_{M,2} = 1.72 \substack{+0.08 \\ -0.08}$, or $5.58\ M_\oplus$) but with a much lower scatter ($\sigma_{M,2} = 0.63 \substack{+0.07 \\ -0.06}$). The low mean mass log-normal, by contrast, has a mean at lower masses ($\mu_{M,1} = 0.60 \substack{+0.49 \\ -0.47}$, or $1.82\ M_\oplus$) with a high scatter ($\sigma_{M,1} = 2.39 \substack{+0.29 \\ -0.24}$. The high mean mass component has about twice the weight of the low mean mass component ($Q_3 = 0.51$ vs $Q_2 = 0.26$). This has the effect of concentrating most of the mass between $1 - 30 M_{\oplus}$, with long tails toward low and high masses. The high mean mass component accounts for the bulk of the super-Earths and mini-Neptunes found by the \textit{Kepler} survey, while the low mean mass component includes planets of all sizes, from sub-Earth sizes to gas giants. Adding this extra component in the mass distribution also has some significant effects in the mass-radius relation, such as shifting the intermediate-mass segment towards lower masses ($\gamma_1 = 0.61 \substack{+0.04 \\ -0.03}$ vs $0.75 \substack{+0.05 \\ -0.05}$, $M_{\text{break},1} = 9.76 \substack{+1.94 \\ -1.43}$ vs $20.0 \substack{+3.4 \\ -3.0}\ M_\oplus$), with lower scatter ($\sigma_1 = 0.23 \substack{+0.05 \\ -0.04}$ vs $0.33 \substack{+0.06 \\ -0.05}$).

We summarize the joint mass-radius-period distribution for model 4 in Figure \ref{Model 4 Main Plot}. We include the resulting 1D mass, period and radius distributions for each mixture, compared to the observed distribution of the sample. We also include a representation of the model in mass-radius space, showing both the mass-radius relation and relative occurrence for all three mixtures, along with the mass-radius measurements included in the sample. We note that these 1D and 2D distributions in Figure \ref{Model 4 Main Plot} are all projections, and do not fully capture the 3D nature of the distribution. While there is no explicit radius or mass dependence in the period distribution, they are not independent in models 2-4 due to the irradiation flux dependence of the envelope mass loss.

\begin{figure*}
    \centering
    \includegraphics[width=0.45\linewidth]{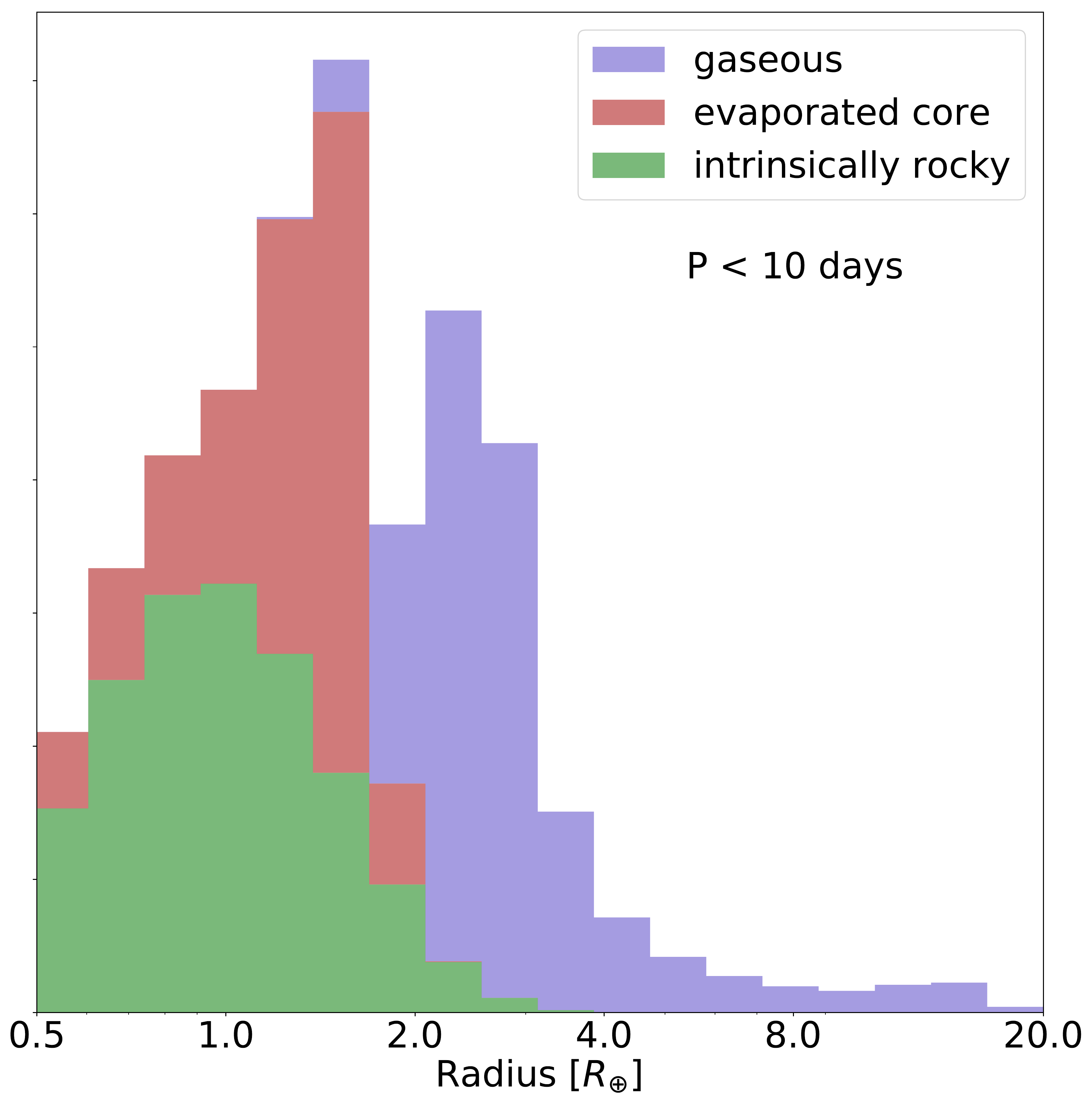}
    \includegraphics[width=0.45\linewidth]{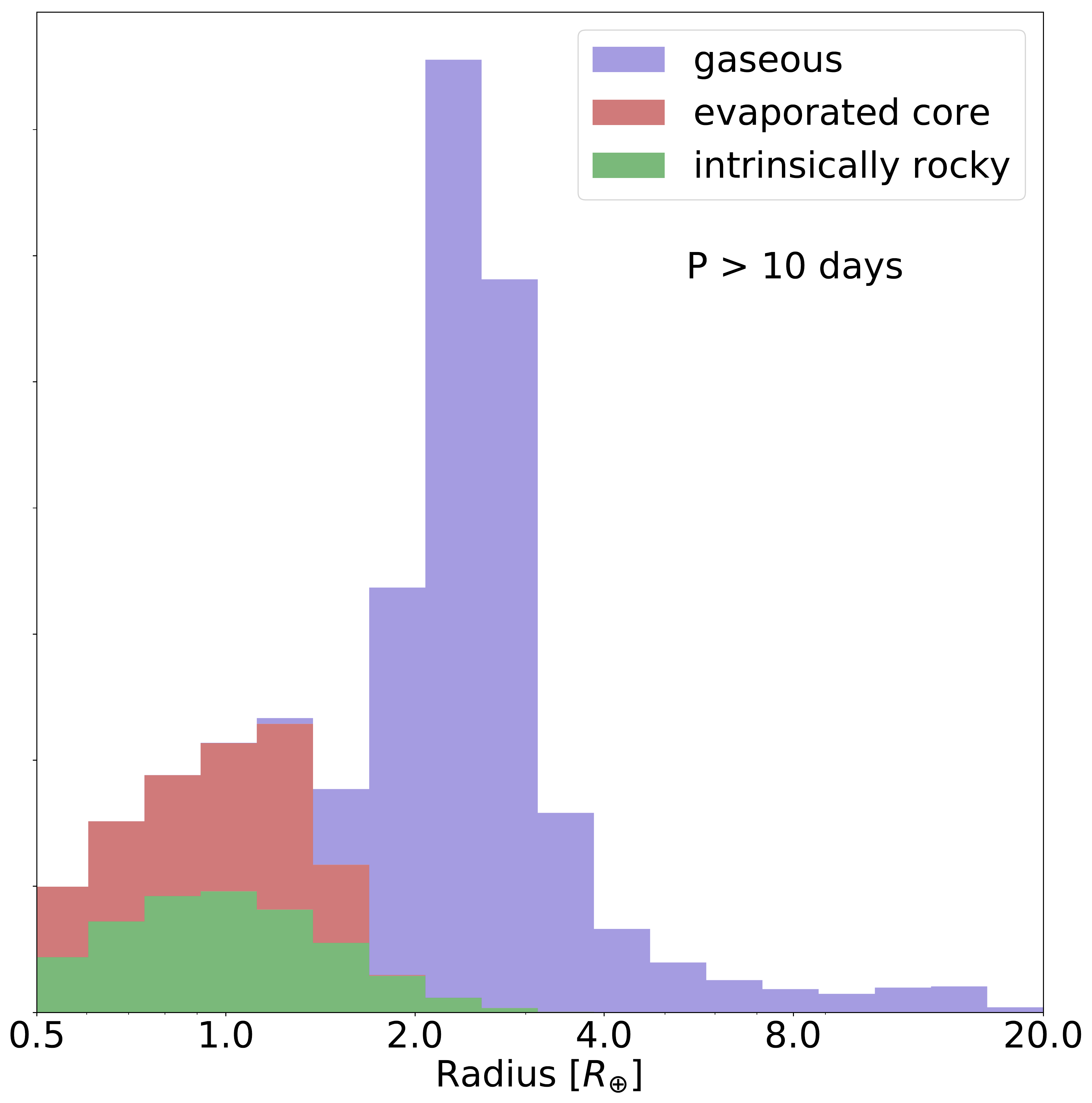}
    \caption{The underlying radius distribution of planets from our model 4, broken down into the three populations of planets: gaseous shown in blue, evaporated cores shown in red, and intrinsically rocky shown in green. On the left we show the radius distribution of planets with short periods ($P < 10 $ days) and on the right we show the radius distribution of planet with long periods ($P > 10 $ days). The three populations are stacked on top of one another, such that the outline shows the overall intrinsic radius distribution. The two radius distributions are normalized separately rather than to each other (there are more planets at $P > 10 $ days than $P < 10 $ days, which is not shown by these distributions)}
    \label{three component radius distribution}
\end{figure*}

\begin{figure}
    \centering
    \includegraphics[width=1\columnwidth]{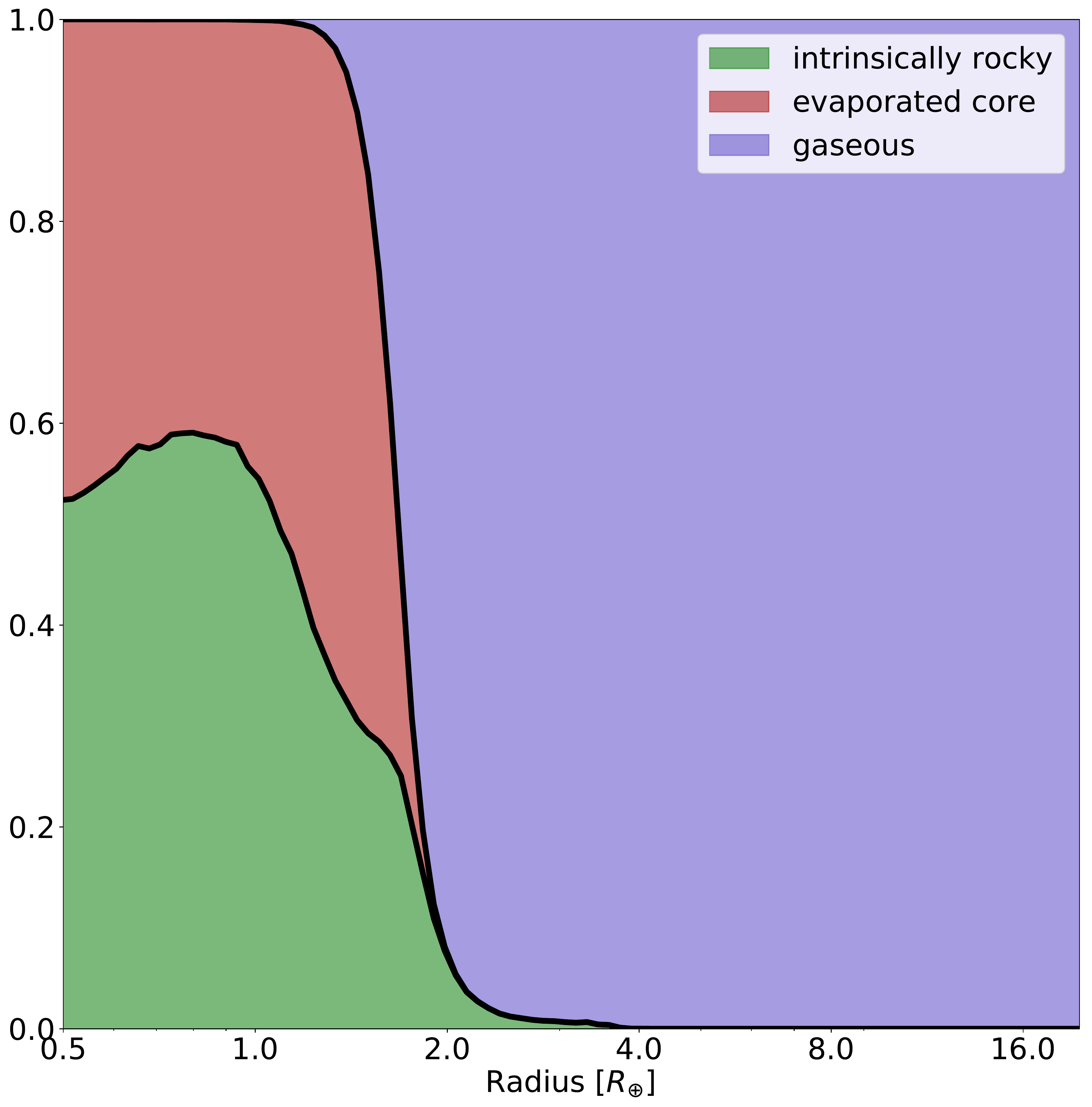}
    \caption{The fraction of planets belonging to each mixture component at a given radius, using our model 4 with three populations of planets: gaseous shown in blue, evaporated cores shown in red, and intrinsically rocky shown in green. }
    \label{fraction of planets in various mixtures}
\end{figure}

We provide an enlarged version of the radius distribution for model 4 in Figure \ref{three component radius distribution} for short period and long period planets, along with the fraction of planets belonging to each mixture as a function of radius in Figure \ref{fraction of planets in various mixtures}. Figure \ref{three component radius distribution} is best compared to Figure 7 in \citet{FultonEt2017ApJ}. We find similar locations for the two peaks, although the gap in between the two is not as deep and even seems to nearly disappear for planets with periods longer than 10 days. We note two possible explanations for this. First, this radius distribution is period dependent. Compared to the radius distribution of planets with $P > 10\ \text{days}$, the radius distribution of planets with $P < 10\ \text{days}$ is shifted towards lower radii, with the $1.3\ R_\oplus$ peak exceeding the height of the $2.4\ R_\oplus$ peak. This is because there are fewer gaseous planets and more evaporated cores and intrinsically rocky planets at short periods. Second, there may not be enough flexibility in our model to allow for a deep gap in the radius distribution, perhaps because of the nature of the mass loss and the smoothness of the log normal distributions. This is further evidenced in Figure \ref{observed radius/period comparison}, where we compare the observed radius distribution of the CKS sample with a simulated observed distribution with our model. The most discrepant bin is the location of the gap, at slightly below $2.0\ R_\oplus$, where our model predicts a significantly lower number of planets.

The transition between the gaseous planets and evaporated cores as shown in both Figures \ref{three component radius distribution} and \ref{fraction of planets in various mixtures} is steep, with significant overlap between the two only in a small range in radius between $1.5$ and $2.0\ R_\oplus$. This is likely mostly due to the envelope mass loss prescription used in the model. We find similar numbers of intrinsically rocky and evaporated core planets, with slightly more intrinsically rocky, especially at low radii ($R < 0.8 R_\oplus$) and high radii ($R > 1.8 R_\oplus$). 

We note that the intrinsically rocky population extends to higher masses and radii than expected. \citet{Rogers2015ApJ} found that most observed planets with $R > 1.6\ R_\oplus$ are not consistent with rocky compositions, yet we find significant amounts of them. For example, as shown in Figure \ref{fraction of planets in various mixtures}, at $R = 2.0 R_\oplus$ about $5\%$ of planets are intrinsically rocky, and there exists a small tail that goes up to $R = 3.0\ R_\oplus$. This is most likely a shortcoming of the model and a data-driven approach: a log-normal mass distribution leads to these high-mass tails, and there is nothing in the model that precludes these planets despite physical expectations for why they should be rare. Simply put, these high-mass rocky planets are massive enough to have accreted significant amounts of gas in the disk during formation \citep{Ikoma&Hori2012ApJ, Bodenheimer&Lissauer2014ApJ}.

We plot the radius-period distributions of the three populations of planets for model 4 in Figure \ref{radius-period occurrence}. The negative slope of the radius valley found by \citet{VanEylenEt2018MNRAS} can be seen by the division between the gaseous planets and the evaporated cores. However, the intrinsically rocky population serves to wash out the radius gap, as the period distribution of the intrinsically rocky population is independent of its radius distribution. While the gap is washed out, adding the intrinsically rocky planets does not eliminate the negative slope of the radius gap, as the fraction of evaporated cores is constrained to be above zero. Therefore we find evidence that there is a population of planets subject to photoevaporation, consistent with \citet{VanEylenEt2018MNRAS}, but the gap that we infer is not as clearly distinguished.

\subsection{Occurrence Rate Comparisons}\label{Occurrence Rate Comparisons}

We compute occurrence rates using each of our four models. Occurrence rates are calculated analytically by integrating Equation \ref{mixture occ rate density} over various ranges in period, mass and radius. We sample over the population-level parameter posteriors in order to incorporate these uncertainties into our occurrence rate estimates. Our reported values and uncertainties are the median and 1-$\sigma$ percentiles from analytically calculating these occurrence rates over $1,000$ posterior samples of the population-level parameters. Our uncertainties do not reflect the uncertainty in the choice of model or parametrization thereof. All reported values are conditional on the model chosen and the data sample used to fit that model.

We compare occurrence rates derived from our models with several values from recent works in Table \ref{occurrence rates table}. We find broad agreement with these previously calculated occurrence rates. Differences likely arise due to the dissimilarities in our data sample and, likely more importantly, model parametrization. Our four 3D joint distribution models are much more complex in their parametrization compared to the common method of constraining the heights of predefined bins. We also note that computing occurrence rates for periods extending past 100 days (the upper limit for our sample and model) involves an extrapolation of our results to these longer periods.

Overall, for the widest radius and period ranges, as in the first two rows of Table \ref{occurrence rates table}, we find that the occurrence rate generally decreases with the added complexities of models 2, 3 and 4. The occurrence rate estimates from each model presented here are consistent within error of the result from \citet{MuldersEt2018AJ}, who modeled the radius-period distribution as independent broken power-laws. The occurrence rates based on our model 1 (with no envelope mass loss) show the best agreement with \citet{MuldersEt2018AJ}, as expected given that model 1's low complexity most closely matches the independent broken power-laws assumed by \citet{MuldersEt2018AJ}. For the mini-Neptune occurrence rates measured by \citet{FultonEt2017ApJ} (rows 3 and 4 in Table \ref{occurrence rates table}), we find that the more complex models predict a higher occurrence rate relative to model 1 and \citet{FultonEt2017ApJ}, despite a lower overall occurrence rate integrated over our full parameter space range. Interestingly, these models also predict a similar occurrence rate for planets between 1.4 and 2.8 $R_{\oplus}$, and planets between 2 and 4 $R_{\oplus}$, whereas \citet{FultonEt2017ApJ} find the former to be slightly more common. Our estimates of the Hot Jupiter occurrence rate are consistent with RV measurements from \citet{MayorEt2011Arxiv}, but are dependent on the model assumed. Our estimates of the Super-Earth occurrence rate are higher than the RV measurements from \citet{HowardEt2010Science}, especially for model 4, which is higher by nearly a factor of 3. We find a rate consistent with previous \textit{Kepler} measurements of the Hot Jupiter occurrence rates with model 1, and a higher rate with models 2 and 3 \citep{PetiguraEt2017AJ}. Interestingly, the more flexible mass distribution used in model 4 decreases this rate in line with model 1 and previous \textit{Kepler} results. 

Our estimates of $\eta_{\oplus}$, as defined by the occurrence rate of planets around GK stars within $20\%$ of $R = 1 R_\oplus$ and $P = 1~\text{year}$, range from 0.08 to 0.008, and are consistent with the wide range of uncertainty reported in \citet{Burke2015ApJ}. However, as with any estimate of $\eta_\oplus$ based on current data, our estimate presented in Table \ref{occurrence rates table} requires extrapolation, and thus is highly model dependent and subject to a large degree of uncertainty not taken into account in the error bars. While our models only constrained the period distribution out to 100 days, calculating $\eta_\oplus$ to within 20\% of a year means that we extrapolate our period distribution to over four times its stated boundary. This assumes the period distribution continues as a power-law from $P_{\text{break}}$ to 438 days. This is a wide range to assume a single power-law; there could easily be another break in the period distribution that we are missing, or perhaps a power-law does not fit this range well at all. Secondly, in this wider period range, any dependence of the period distribution on radius or mass is likely to be more significant. While our modeling of three mixture populations allows for correlations between mass or radius and period (e.g. fewer low-density planets at short orbital periods due to hydrodynamic envelope mass loss), there can be additional correlations between these quantities within a mixture that we are not accounting for. Thus, care must be taken to avoid overinterpretation of these (and all other) extrapolated $\eta_{\oplus}$ estimates. 

Despite these caveats, there are important lessons to take away in our extrapolation of $\eta_\oplus$. We find that our estimates vary by nearly an order of magnitude depending on the choice of model: $\eta_\oplus$ calculated with model 1 is consistent with \citet{Burke2015ApJ}, whereas models 2-4 find a much lower $\eta_\oplus$. In model 1, the period distribution is completely independent of the radius and mass distributions, so rocky planets and gaseous planets extrapolate similarly to longer periods, and the ratio between the two remains the same at short orbits versus longer orbits. When we introduce envelope mass loss with model 2, as we go out to longer orbital periods the proportion of gaseous planets relative to rocky planets increases, as planets at these large distances from their star can more easily retain their envelopes. Accordingly, the $\eta_\oplus$ estimate drops by nearly an order of magnitude, as planets with H/He envelopes typically have radii larger than 1.2 $R_\oplus$.

\begin{figure}
    \centering
    \includegraphics[width=1\columnwidth]{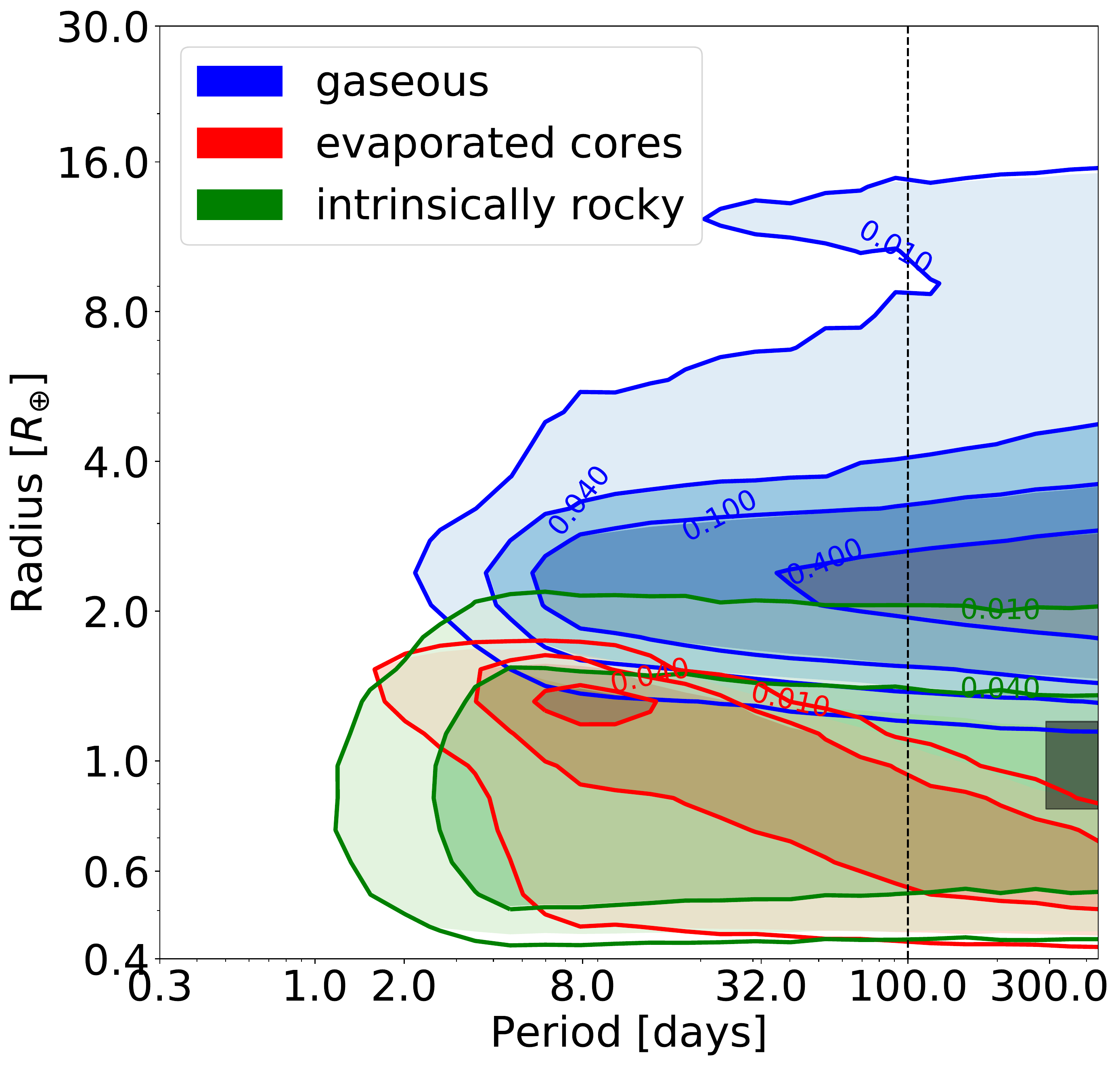}
    \caption{Contours of occurrence rates in radius-period space for the gaseous, evaporated core, and intrinsically rocky populations from model 4. The vertical line at 100 days shows the boundary to which we originally fitted our models, and the grey box shows the region in which we calculate $\eta_\oplus$.}
    \label{radius-period occurrence}
\end{figure}

Though the overall extrapolated $\eta_\oplus$ value is similar for models 2-4, the mass (and hence radius) distribution of the extrapolated planet populations at $\approx 1 \text{AU}$ are very different. To illustrate this, we refer to Figure \ref{radius-period occurrence}, where we highlight the bounds in which we calculate $\eta_\oplus$. For model 2, planets that fall in the $\eta_\oplus$ bin are split between the gaseous and evaporated core populations (see section \ref{Caveats} for further discussion). The inclusion of an intrinsically rocky population in models 3-4 allows for more massive rocky planets at these longer orbital periods, compared to the evaporated core population. Accordingly, as shown in Figure \ref{radius-period occurrence}, the intrinsically rocky planets are much more evenly distributed in radius inside the bin, whereas the evaporated core planets are preferentially lower radii.

In our calculation of $\eta_\oplus$, we used the same boundaries in radius and period as \citet{Burke2015ApJ}, to make the comparison direct. However, we are neglecting the true power of the full mass-radius-period distribution in doing so. As mentioned above, in our calculation of $\eta_\oplus$ we were including planets of a wide range of densities in this bin. By adding constraints on the mass range in calculating $\eta_\oplus$, we can ensure the planets are more Earth-like. We add a bound of a factor of 2 on either side of an Earth mass (0.5 - 2 $M_\oplus$) and recalculate $\eta_\oplus$. We find that the numbers drop even more, with $0.055 \substack{+0.011 \\ -0.009}$ for model 1, and $0.005 \substack{+0.003 \\ -0.002}$, $0.006 \substack{+0.004 \\ -0.003}$, and $0.008 \substack{+0.006 \\ -0.004}$ for models 2, 3 and 4, respectively. We note that with these bounds, model 3 is higher than model 2, in contrast to when calculated with no mass bounds. Clearly, this wide range of answers should give even more pause when considering any extrapolated $\eta_\oplus$ calculations for planets around GK dwarfs at face value.

The comparisons in Table \ref{occurrence rates table} were all derived using \textit{Kepler} data, with the exception of the comparison to the RV-measured Hot Jupiter occurrence rate from \citet{MayorEt2011Arxiv} as well as the RV-measured Super-Earth occurrence rate from \citet{HowardEt2010Science}. With the full mass-radius-period distribution modeled here, comparisons with surveys that measure planet mass, such as RV and microlensing surveys, are more direct and straightforward than if we were to simply model the radius-period distribution. However, differences in period and host star distributions, among other effects, still make direct comparisons with these surveys difficult. Nonetheless, our occurrence rates derived here are more generalizable to a wider range of parameter space, and our comparisons demonstrate their validity by showing consistency with prior results. The model dependence of our results, and to a further extent all occurrence rate calculations, must be kept in mind to avoid overinterpretation.

\setlength\extrarowheight{3pt}
\begin{table*}
    \centering
    \begin{tabular}{c c c c c c c}
    \hline
    Variable Range & Model 1 & Model 2 & Model 3 & Model 4 & Literature Value & Reference\\
    \hline\hline
    $0.4 < \frac{R}{R_{\oplus}} < 30$ , $0.3 < \frac{P}{\text{1 day}} < 100$ & $1.36 \substack{+0.07 \\ -0.06}$ & $1.31 \substack{+0.06 \\ -0.06}$ & $1.10 \substack{+0.05 \\ -0.05}$ & $1.12 \substack{+0.06 \\ -0.05}$ & - & - \\
    $0.5 < \frac{R}{R_{\oplus}} < 6$ , $2 < \frac{P}{\text{1 day}} < 400$ & $2.28 \substack{+0.22 \\ -0.18}$ & $1.89 \substack{+0.18 \\ -0.13}$ & $1.70 \substack{+0.18 \\ -0.17}$ & $1.70 \substack{+0.21 \\ -0.17}$ & $2.4 \substack{+0.5 \\ -0.5}$ & \citet{MuldersEt2018AJ} \\
    $1.4 < \frac{R}{R_{\oplus}} < 2.8$ , $0.3 < \frac{P}{\text{1 day}} < 100$ & $0.43 \substack{+0.02 \\ -0.02}$ & $0.47 \substack{+0.03 \\ -0.03}$ & $0.48 \substack{+0.04 \\ -0.04}$ & $0.51 \substack{+0.05 \\ -0.05}$ & $0.43 \substack{+0.02 \\ -0.02}$ & \citet{FultonEt2017ApJ} \\
    $2 < \frac{R}{R_{\oplus}} < 4$ , $0.3 < \frac{P}{\text{1 day}} < 100$ & $0.30 \substack{+0.01 \\ -0.01}$ & $0.45 \substack{+0.02 \\ -0.02}$ & $0.46 \substack{+0.03 \\ -0.03}$ & $0.49 \substack{+0.04 \\ -0.04}$ & $0.37 \substack{+0.02 \\ -0.02}$ & \citet{FultonEt2017ApJ} \\
    $50 < \frac{M}{M_{\oplus}} < 10000$ , $0.3 < \frac{P}{\text{1 day}} < 11$ & $0.005 \substack{+0.001 \\ -0.001}$ & $0.010 \substack{+0.002 \\ -0.001}$ & $0.009 \substack{+0.001 \\ -0.001}$ & $0.005 \substack{+0.001 \\ -0.001}$ & $0.009 \substack{+0.004 \\ -0.004}$ & \citet{MayorEt2011Arxiv} \\
    $3 < \frac{M}{M_{\oplus}} < 10$ , $0.3 < \frac{P}{\text{1 day}} < 50$ & $0.19 \substack{+0.01 \\ -0.01}$ & $0.22 \substack{+0.01 \\ -0.01}$ & $0.22 \substack{+0.02 \\ -0.02}$ & $0.32 \substack{+0.03 \\ -0.03}$ & $0.12 \substack{+0.04 \\ -0.04}$ & \citet{HowardEt2010Science} \\
    $0.8 < \frac{R}{R_{\oplus}} < 1.2$ , $292 < \frac{P}{\text{1 day}} < 438$ & $0.076 \substack{+0.016 \\ -0.011}$ & $0.008 \substack{+0.003 \\ -0.002}$ & $0.008 \substack{+0.004 \\ -0.003}$ & $0.009 \substack{+0.007 \\ -0.004}$ & $0.1 \substack{+1.9 \\ -0.1}$ & \citet{Burke2015ApJ} \\
    \begin{tabular}{c} $0.8 < \frac{R}{R_\oplus} < 1.2$, $292 < \frac{P}{\text{1 day}} < 438$, \\ $0.5 < \frac{M}{M_{\oplus}} < 2.0$ \end{tabular} & $0.055 \substack{+0.011 \\ -0.009}$ & $0.005 \substack{+0.003 \\ -0.002}$ & $0.006 \substack{+0.004 \\ -0.003}$ & $0.008 \substack{+0.006 \\ -0.004}$ & - & - \\
    \hline
    \end{tabular}
    \caption{The occurrence rate within different intervals in mass, period and radius for each of the four models, along with a comparison to previous works.}
    \label{occurrence rates table}
\end{table*}

\subsection{Model Selection}\label{Model Selection}

In the preceding sections we have presented four viable models for a joint mass-radius-period distribution, with increasing complexity. While the additional complexities of the latter models are motivated by physical processes and planet formation theory, we would like an objective measure of whether or not these additional complexities are justified, or even necessitated, by the current data. In order to assess this, we must measure each model's predictive accuracy, or its ability to generalize to an independent sample of planets, while accounting for problems such as overfitting. 

Cross-validation is a class of model selection techniques which uses the sample of interest in order to estimate the out-of-sample predictive accuracy \citep{HastieSpringer2001}. The most comprehensive cross-validation is leave-one-out cross-validation (LOO), where for each data point the model is rerun leaving out that point, and then the likelihood of the left-out point is calculated using that model fit. Given that our dataset contains 1130 planets, this would require 1130 model runs and thus is computationally prohibitive. There are several approximations to LOO that don't necessitate additional model runs, such as Watanabe-Akaike information criterion \citep{Watanabe10}, used in previous analyses \citep{Neil&Rogers2018ApJ}, but diagnostics (e.g. the variance over the posterior samples of the log-likelihood of a given planet) indicate that these are invalid for our current models, possibly due to their three-dimensional nature in period, mass and radius \citep{VehtariEt2017}. In the field of exoplanets, cross-validation has been used for model selection in predicting additional transiting planets in \textit{TESS} systems \citep{Kipping&LamMNRAS2017} as well as improving transit classification \citep{AnsdellEt2018ApJL}. We turn to K-fold cross-validation as our model selection method of choice.

We choose 10 folds for our cross-validation as a compromise between computational time and robustness. We divide the planet sample into 10 subsets, separately dividing the planets with mass measurements and those without mass measurements to ensure the ratio between the two is consistent between subsets. We then create our 10 new planet samples by excluding the planets from one of the 10 subsets. We fit our four models on each of these 10 samples, and calculate the expected log predictive density (elpd) for each planet using the model fit in which the planet was excluded:

\begin{equation} \label{log predictive density}
\begin{split}
    \widehat{\text{elpd}}_k = & \log p(x_k|\left\{x_{(-j)}\right\}) \\
    \log p(x_k|\left\{x_{(-j)}\right\}) = & \log \left( \frac{1}{S} \sum_{s=1}^{S} p(x_k|\theta^{j,s}) \right)\
\end{split}
\end{equation}

\noindent where the data for a given planet $k$ is summarized by $x_k$; for a given subset $j$ of the planet sample, the sample excluding that subset is $x_{(-j)}$ and the resulting model fit is $\theta^j$, and we can use a total of $S$ draws from our model posteriors to summarize our posterior distribution \citep{VehtariEt2017}. For our joint distribution, generalizing to mixture models, this probability turns into:

\begin{equation} \label{planet likelihood}
\begin{split}
    p(x_k|\theta) = & \ \sum^{N_{\text{mix}}}_{q=0} p(q | M_{obs,k}, P_k, \theta) (P_k | \theta, q) \\
    & \ \cdot p(R_{obs,k} | M_{obs,k}, \theta, q) p(M_{obs,k} | \theta, q)
\end{split}
\end{equation}

\noindent for planets with mass measurements, where $q$ gives the index for a given mixture component. For planets without mass measurements we remove the last term and instead marginalize over the mass distribution for the radius term. Our total expected log predictive density (elpd) for the entire model is then:

\begin{equation} \label{elpd}
    \widehat{\text{elpd}} = \sum_{k=1}^K \log p(x_k|x_{(-j)})
\end{equation}

\noindent where we are summing over each planet's contribution from Equations (\ref{log predictive density}), (\ref{planet likelihood}), with $K$ representing the number of planets. We can then compare the expected log predictive density between two models directly, with the standard error (se) of the difference as follows:

\begin{equation} \label{elpd error}
    \text{se} (\widehat{\text{elpd}}^A - \widehat{\text{elpd}}^B) = \sqrt{N V_{k=1}^K (\widehat{\text{elpd}}_k^A - \widehat{\text{elpd}}_k^B)}
\end{equation}

\noindent where $A$ and $B$ represent two different models, and $V$ represents the sample variance (in this case the variance of the differences).

\begin{table}
\centering
\begin{tabular}{c c c} 
 \hline
 Model & $\widehat{\text{elpd}} - \widehat{\text{elpd}}_\text{M1}$ \\
 \hline\hline
 Model 2 & -257 $\pm$ 18 \\
 Model 3 & -380 $\pm$ 19 \\
 Model 4 & -376 $\pm$ 20 \\
 \hline
\end{tabular}
\caption{Difference in expected log predictive density for each model compared to model 1, along with the error in the difference. Negative numbers favor the model in question over model 1. We find that each model with envelope mass loss is strongly preferred over the model without. Models 3 and 4, which include multiple populations of rocky planets, are preferred over model 2, which only has evaporated cores. From these results, Models 3 and 4 are statistically indistinguishable from each other and one is not preferred over the other.}
\label{table:crossval}
\end{table}

We present our results in Table \ref{table:crossval}. The results presented are in terms of the difference in expected log predictive density between the given model and model 1, which has no envelope mass loss. Negative numbers favor the model in question. We find that each model with envelope mass loss (models 2, 3 and 4) is strongly preferred over the basic model with a single population of planets and no envelope mass loss (model 1). Furthermore, models 3 and 4, which include a second, independent population of rocky planets, are both preferred over model 2. The difference in the expected log predictive density between models 3 and 4 is only 4, with an uncertainty in the difference of 11; thus, these results do not strongly prefer either model 3 or 4 over the other. 

Given these results, we can confidently suggest that we are justified in the additional complexities introduced in the latter models. Not only are these complexities easily physically motivated, but the cross-validation shows that these models better predict out-of-sample data. One of the main purposes of this cross-validation is to make sure we are not overfitting the data and fitting peculiarities in this particular sample that do not generalize to the exoplanet population as a whole. That seems to not be the case. While 25 population-level parameters for model 4 seems like a large number, our dataset contains 1130 planets, with each planet having a radius and period measurement, and 53 of those planets having a mass measurement, for a total of 2313 measurements. This is nearly 100 times the number of parameters. Further complexity than what we have modeled here may even yield further improvement in the expected log predictive density, especially in the radius and period distributions, for which we have plenty of data.

\begin{figure*}
    \centering
    \includegraphics[width=0.45\linewidth]{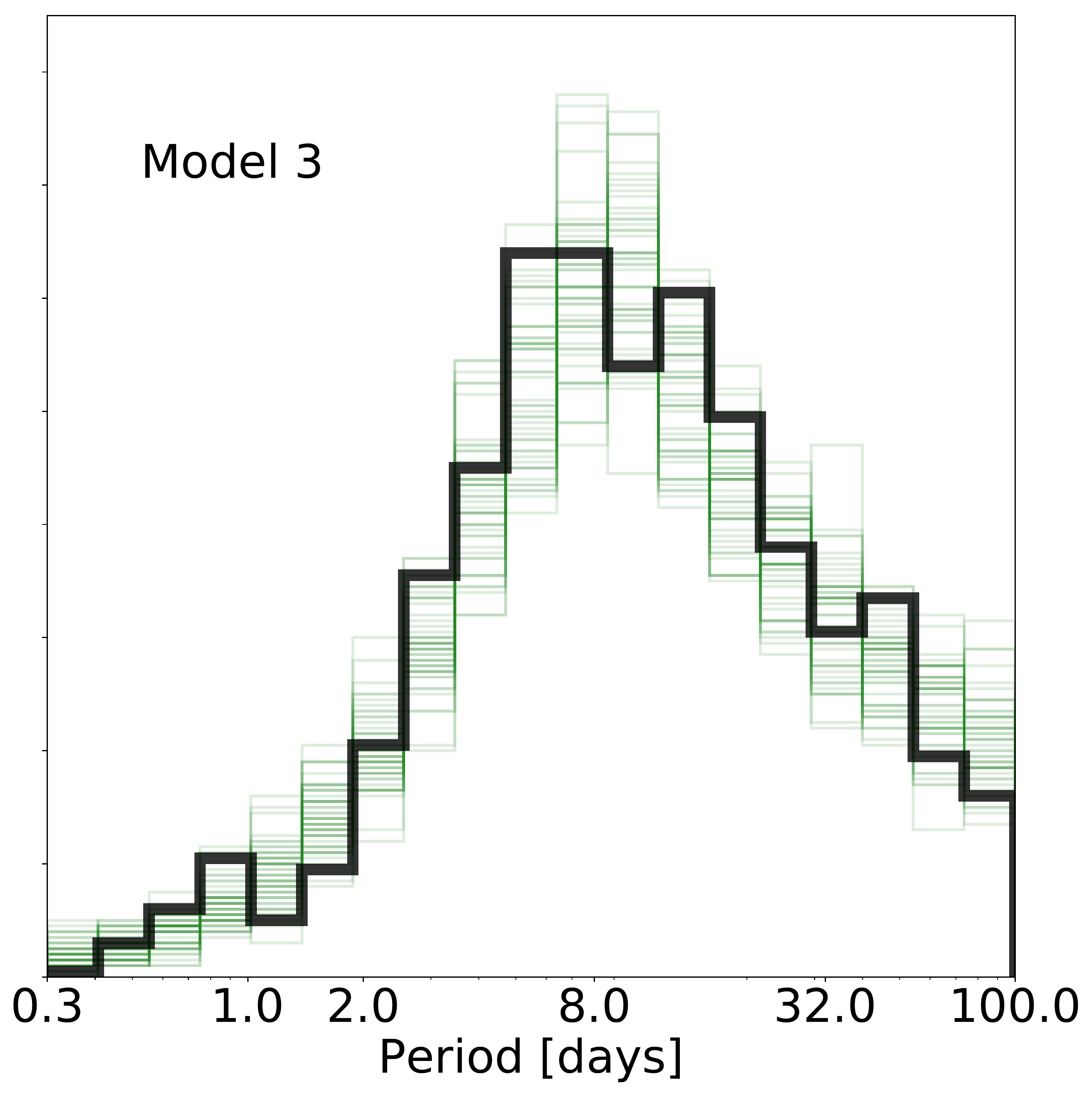}
    \includegraphics[width=0.45\linewidth]{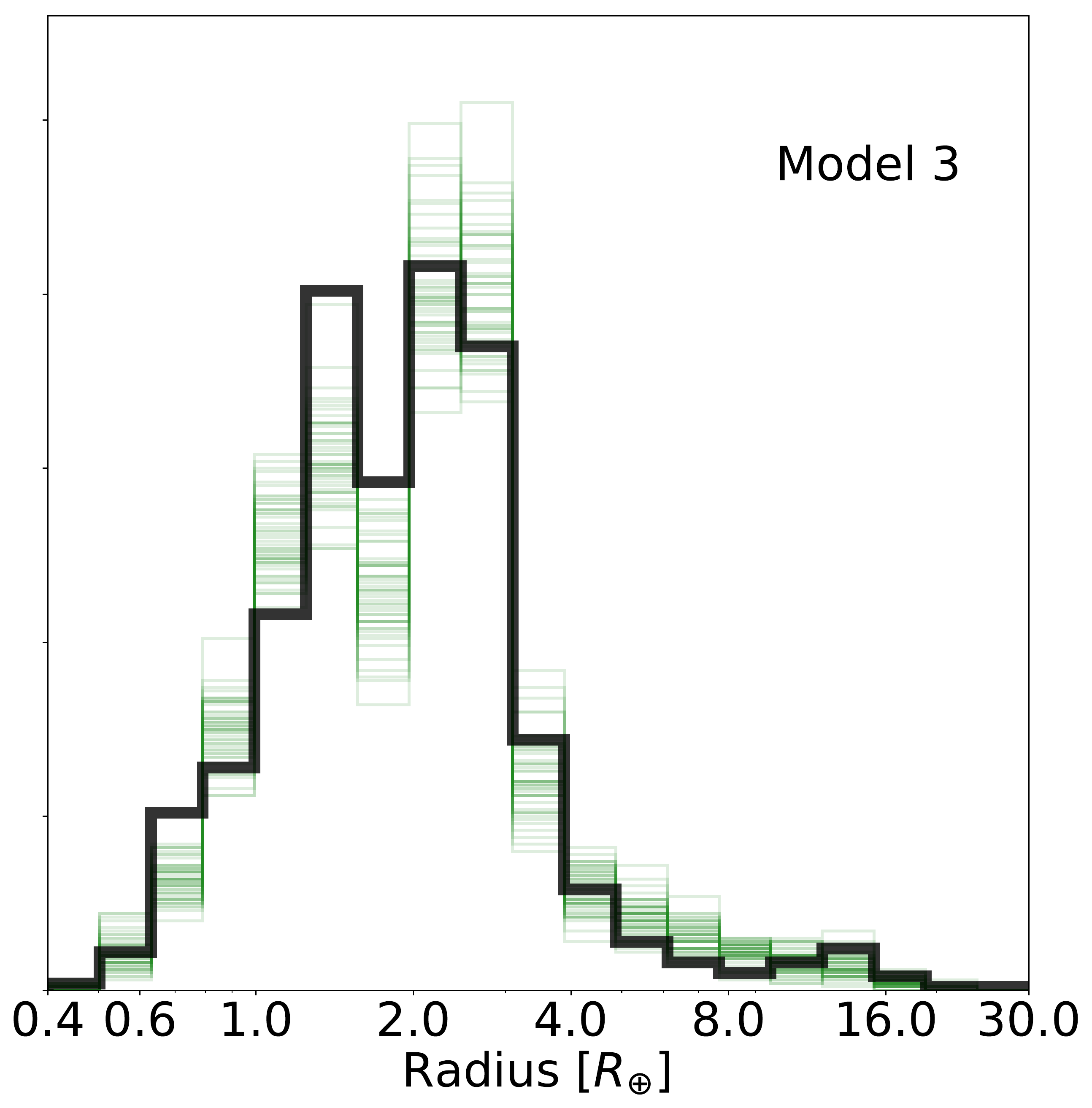} \\
    \includegraphics[width=0.45\linewidth]{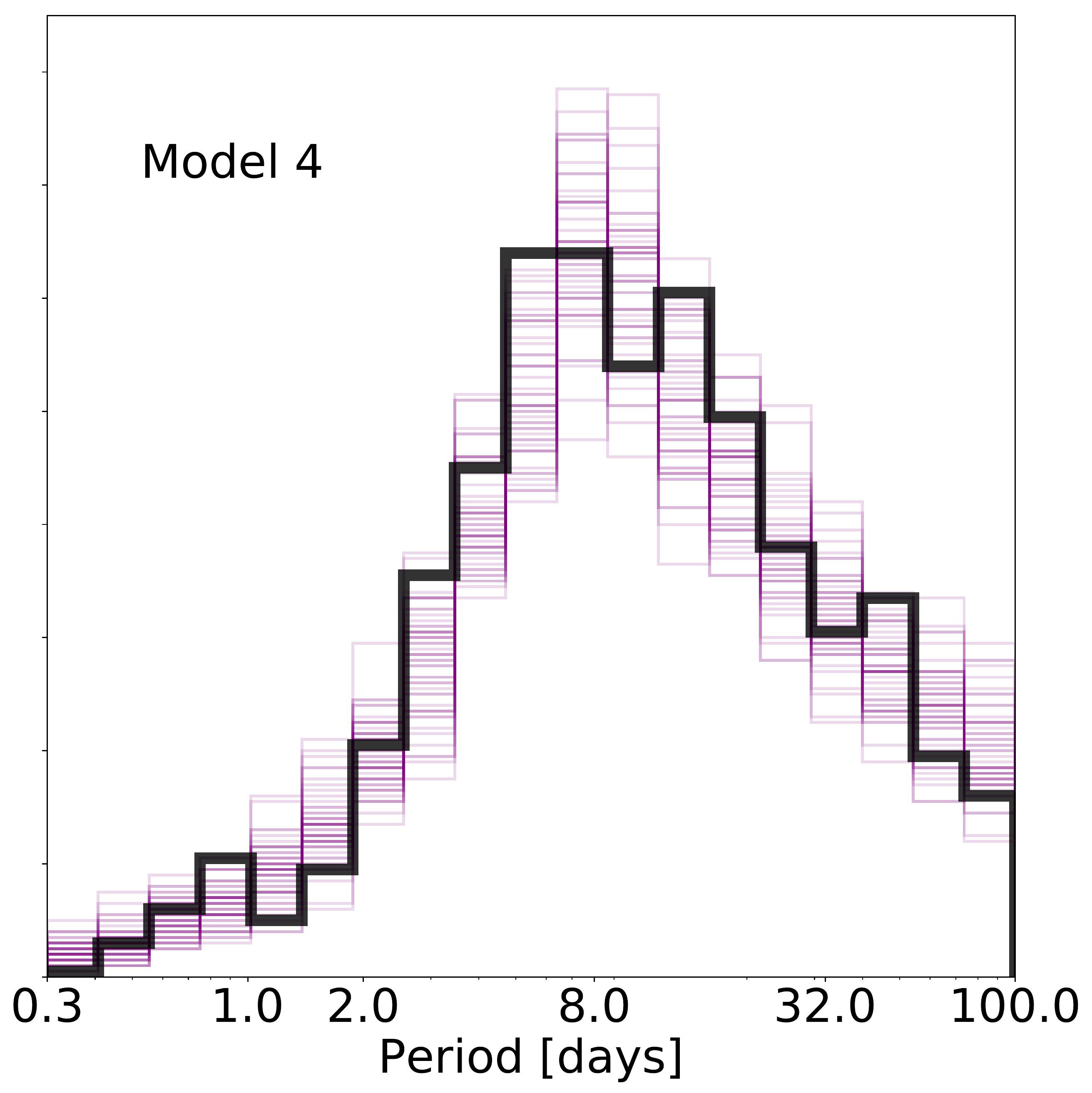}
    \includegraphics[width=0.45\linewidth]{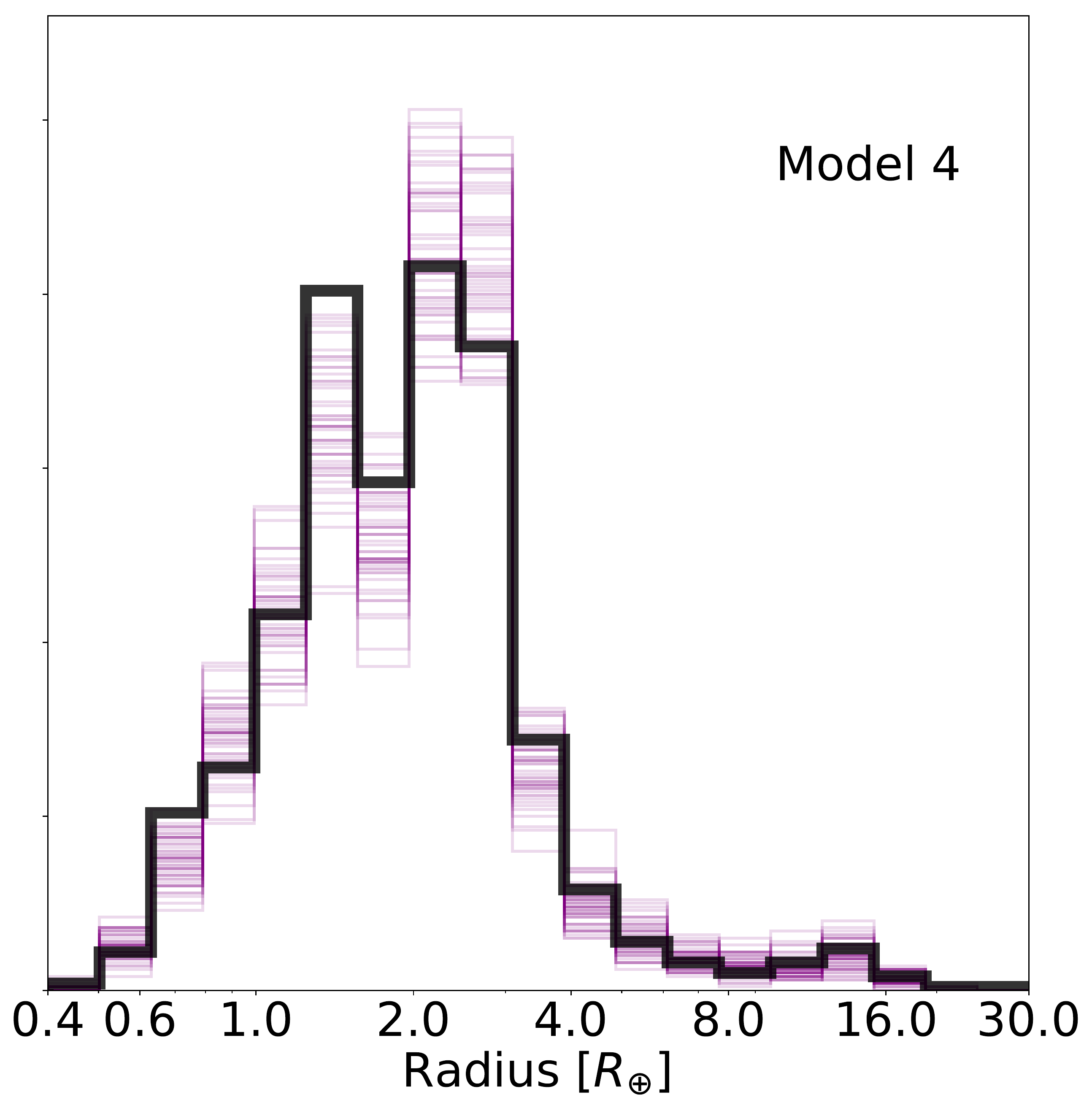}
    \caption{The observed period (left) and radius (right) distributions of the \textit{Kepler} sample, shown in black, compared to the simulated observed distributions using our model 3 with three planet populations (top) shown in green, and model 4 with an extra gaseous log-normal mass component (bottom), shown in purple. We draw planets from our joint MRP distribution and then run them through the \textit{Kepler} detection efficiency to create simulated observed catalogs of planets. We repeat this 50 times using draws from our model posteriors to show the spread in the simulated distributions.}
    \label{observed radius/period comparison}
\end{figure*}

The cross-validation results do not strongly prefer model 3 over model 4, or vice versa. However, there are other pieces of evidence we can turn to in assessing these models. The difference between the observed radius distribution of the real dataset and a simulated dataset generated with the model fit, as shown in Figure \ref{observed radius/period comparison} for models 3 and 4, shows that model 4 better predicts the radius distribution at the highest radii (above $4 R_{\oplus}$), due to the longer tail towards high masses in the mass distribution. Model 4 also better predicts the radius distribution at small radii (below $1.5 R_{\oplus}$). These features may not show up in the cross-validation results because the expected log predictive density is a sum of all the log likelihoods for each planet, and giant planets as well as very small planets represent a small fraction of the overall dataset. Model 4 also has a smaller tail towards high masses for planets of about $2 R_{\oplus}$. This is evident in Figure \ref{mass predictions}, where for Kepler-60d the mass prediction using model 4 falls off more steeply past $15 M_{\oplus}$ than for either model 2 or 3. These effects are relatively small, and it is reasonable that the cross-validation can not distinguish between model 3 and 4. 

\subsection{Mass/Radius Predictions}\label{Mass/Radius Predictions}

With a joint mass-radius-period distribution, we can predict planet masses from radii and vice versa. Compared to predicting masses or radii using a mass-radius relationship alone, our models have several benefits. First, the fact that we are constraining the underlying distributions, rather than just the relationship between mass and radius, allows us to weight the observation accordingly. For example, if a planet has a measured mass of $20 M_{\oplus}$ with a Gaussian error of $5 M_{\oplus}$, the low end of the mass range should be more strongly weighted than the high end, since in this case planets in this less massive range are more common. This would lead to a smaller radius prediction than if one had used a mass-radius relation in isolation. Secondly, the additional complexities that we model here, including envelope photo-evaporation and mixtures of distinct sub-populations of planets, are physically motivated and would be difficult to model with a mass-radius relation alone. This allows for situations where the resulting radius or mass prediction can be bimodal, depending on which mixture the planet falls into. The predictions for two similar size planets can also differ, if their periods are dissimilar.

\begin{figure*}
    \centering
    \includegraphics[width=0.45\linewidth]{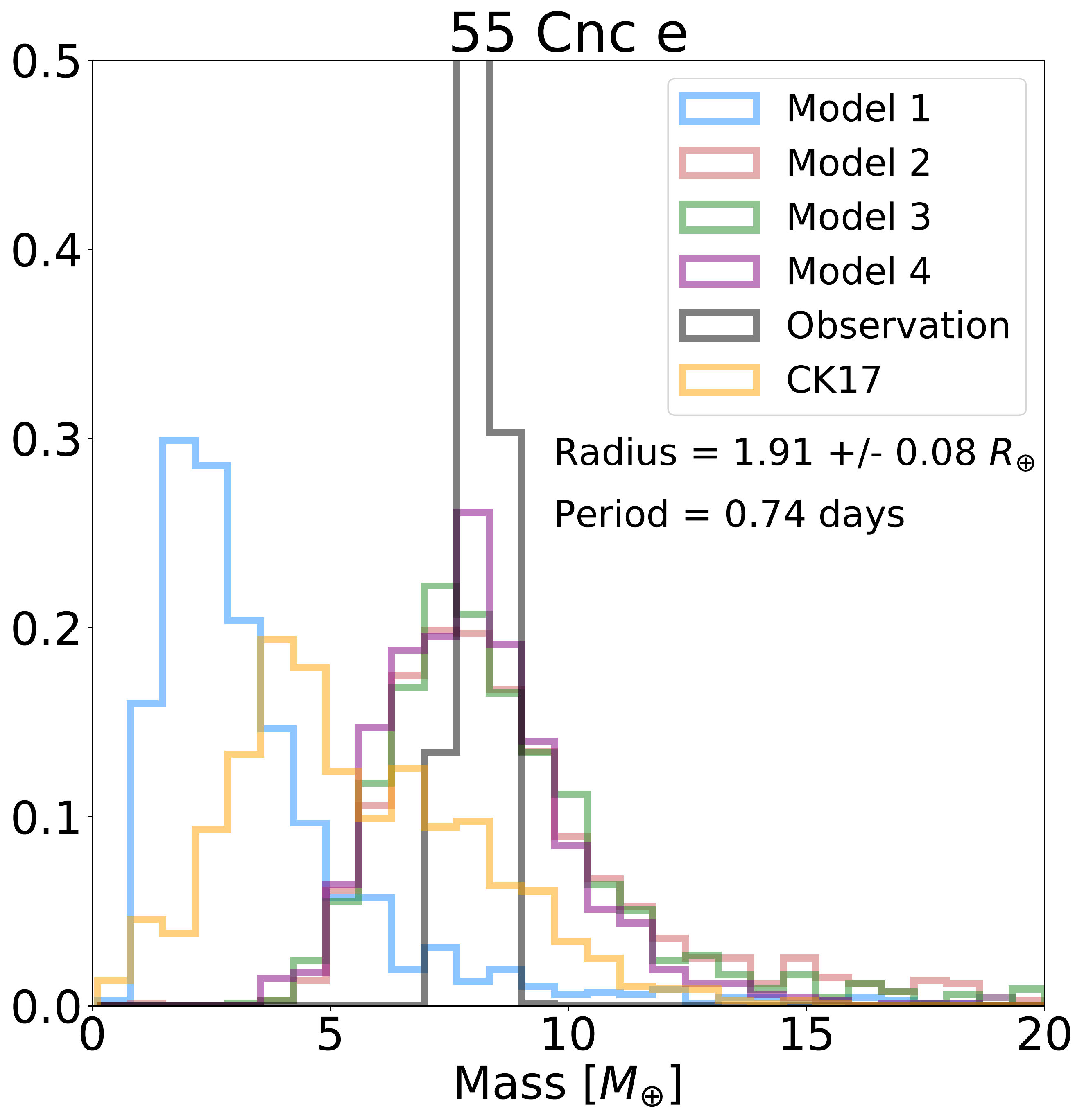}
    \includegraphics[width=0.45\linewidth]{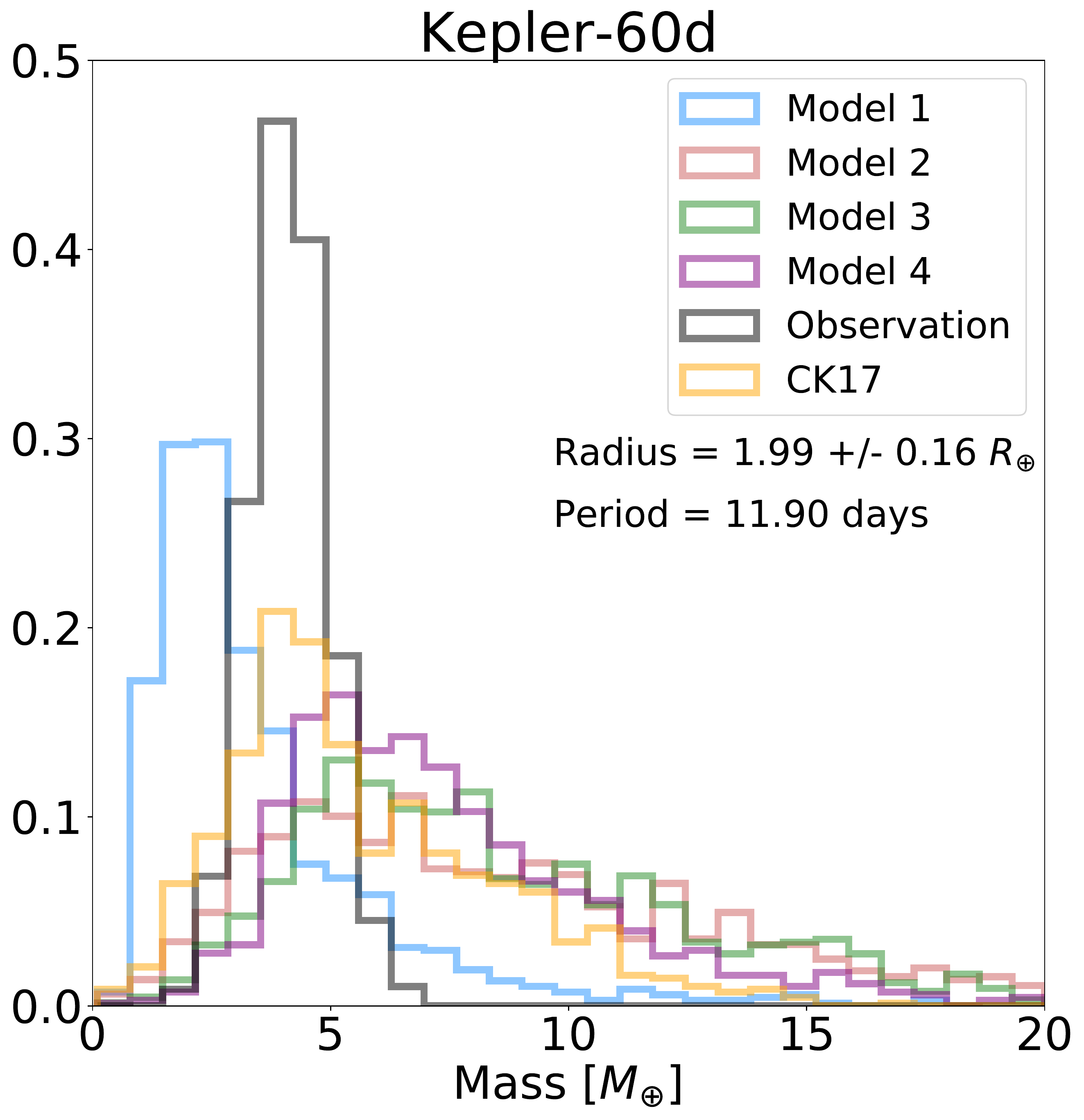}
    \caption{Mass prediction using our posteriors from each of our four models compared to the mass prediction from \citet{ChenKipping2017ApJ} as well as the actual mass measurement, for two different planets with similar radii but dissimilar compositions. Both 55 Cnc-e and Kepler-60d have radii close to $2 R_{\oplus}$, but 55 Cnc-e has a mass of $8.1 \pm 0.3 M_{\oplus}$ and a period of $0.74$ days, and Kepler 60-d has a mass of $4.2 \pm 0.8 M_{\oplus}$ and a period of $11.9$ days.}
    \label{mass predictions}
\end{figure*}

We illustrate the latter effect in Figure \ref{mass predictions}, where we show mass predictions using our models for two different planets, 55 Cnc e and Kepler-60d. These two planets have similar radii (1.91 vs 1.99 $R_{\oplus}$, respectively), but dissimilar periods (0.74 vs 11.9 days). For our models with envelope mass loss (models 2-4), predicting masses for these two planets leads to two substantially different predictions. 55 Cnc e is on a very short orbit, and thus it is likely that the planet has lost its envelope due to hydrodynamic mass loss, and should have a rocky composition (in the context of our model since we are not including ices). The mass predictions for this planet using the models that take into account envelope mass loss consequently predict a peak mass of around $8 M_{\oplus}$, in line with the RV observations, with a sharp cutoff below $4 M_{\oplus}$. On the other hand, Kepler 60-d is on a longer orbit and may have retained a gaseous envelope. The mass predictions for this planet using the envelope mass loss models thus allow for lower mass predictions and peak at around $5 M_{\oplus}$, again in line with the TTV mass measurement.

55 Cnc e and Kepler-60d are similar sizes, but their disparate periods allow for different mass predictions when envelope mass loss is taken into account. By contrast, estimates using the probabilistic broken powerlaw mass-radius relation of CK17 or our model 1, which don't take into account envelope mass loss or the existence of multiple exoplanet populations, predict nearly the same mass for both planets. Note that we are not assessing the generalized accuracy of our mass or radius predictions. Properly testing this generalized accuracy would require not just two test cases, but an independent sample of planets with mass and radius measurements that were not included in fitting the model. Our example in Figure \ref{mass predictions} serves to illustrate one of the consequences of using our models for predicting planet masses.

\section{Discussion}\label{Discussion}

\subsection{Mixture Labels}\label{Mixture Labels}

\begin{figure}
    \centering
    \includegraphics[width=1\columnwidth]{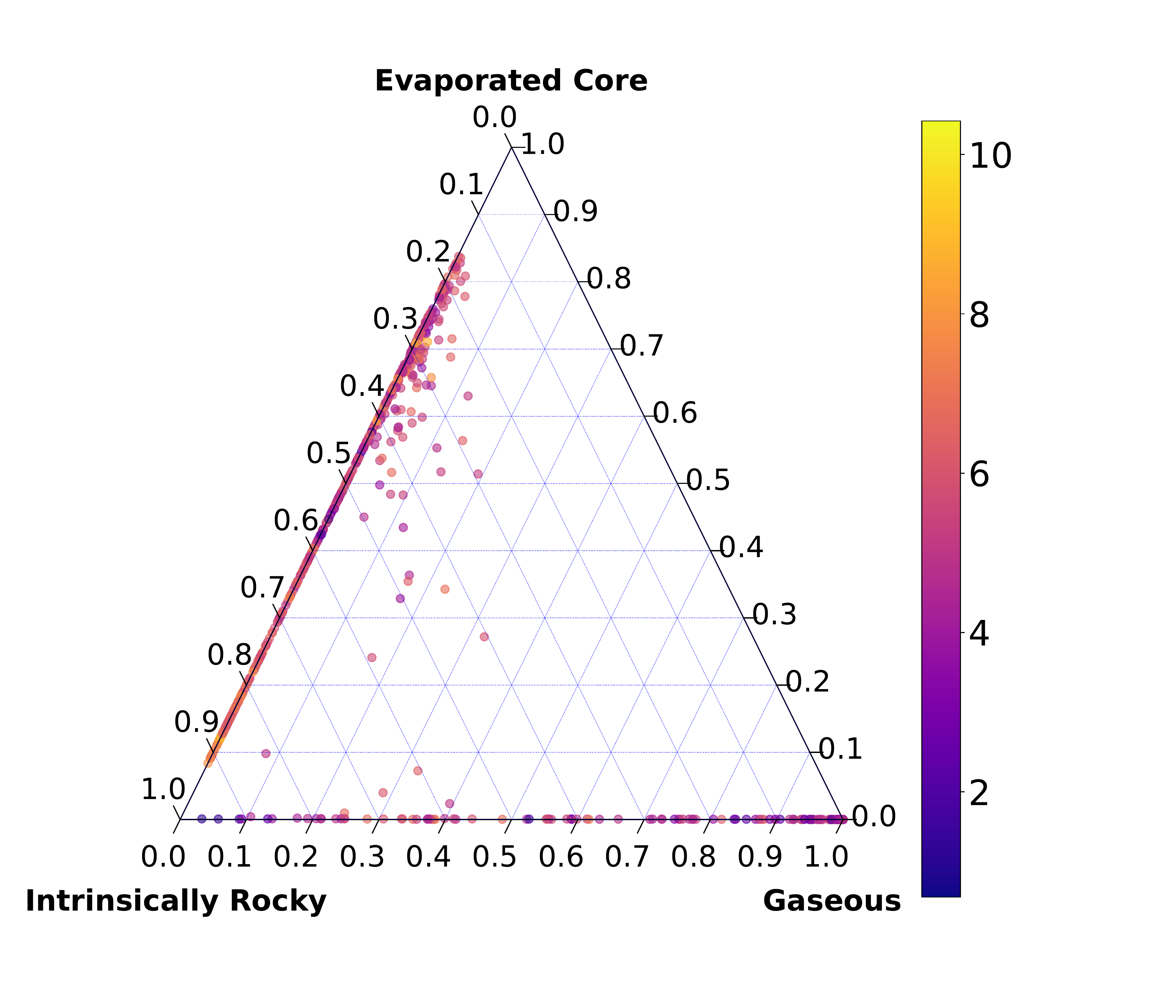}
    \caption{Ternary plot showing the distribution of the retrieved probabilities of the three mixture components for the 1130 planets in our sample from model 4. Each point represents a planet and are colored by incident flux (scaled to the insolation flux from the Sun at Earth's orbit). The closer a planet is to a labeled corner, the higher the probability of that planet belonging to that mixture. The axis for each mixture component is located in the clockwise direction from the mixture component's respective corner.}
    \label{ternary plots}
\end{figure}

While we have given each mixture component physically intuitive labels, i.e. ``gaseous", ``evaporated cores" and ``intrinsically rocky", we note that these labels do not necessarily reflect the formation history for each individual planet. These mixture components are an abstraction that we have used to approximate the various physical populations of planets. For instance, given the large amount of overlap between the evaporated core and intrinsically rocky populations, some planets that fall under one label could be the other, and the intrinsically rocky population could be serving to give more flexibility to the period and mass distributions of the evaporated cores. As a further example, since the mass-radius relation is described by a normal distribution at a given mass, the low radius tail of gaseous planets at low masses would have compositions consistent with rocky, and would not be considered gaseous planets. These high density ``gaseous" planets fall into the estimate for $\eta_\oplus$, as described in section \ref{Occurrence Rate Comparisons}.

In order to visualize the distribution of these labels across our planet sample, we present a ternary plot in Figure \ref{ternary plots} which shows the membership probabilities (gaseous vs evaporated core vs intrinsically rocky), as well as incident flux, for every planet in our sample. We find that most planets in the sample fall along two axes: the bottom axis, which is split between intrinsically rocky and gaseous planets, and the left axis, which is split between evaporated core and intrinsically rocky planets. There is also a substantial number of planets with low (less than 10\%) gaseous probability that fall somewhere between the evaporated core and intrinsically rocky axis. Notably, there is a significant lack of planets with roughly even probabilities between the three mixtures, and a significant lack of planets along the right axis, with close to zero probability of being intrinsically rocky and split between evaporated core and gaseous. There are also no planets we can definitively say are intrinsically rocky and not evaporated cores, based on our model. We take this as evidence that the intrinsically rocky population is adding flexibility to the mass and period distributions of the evaporated cores, and cannot necessarily prove the existence of this separate population based on our data sample.

As a further test of the validity of including this intrinsically rocky population, we tested two additional models, as modifications of our four models presented in this work. Our first alternate model adds another component to the intrinsically gaseous population in model 2, with a separate period and mass distribution. This can be seen as an alternative to model 3, where we add flexibility to the intrinsically gaseous population rather than creating an intrinsically rocky population, so we label it ``model 3b". The second modified model adds a separate period distribution to the second component of the mass distribution for the intrinsically gaseous population in model 4, so we label it ``model 4b". The aim of these modified models is to test whether the additional flexibility to the intrinsically gaseous population eliminates the need for the intrinsically rocky population. We find this to not be the case. Using the 10-fold cross-validation in Section \ref{Model Selection}, we find that model 3 is significantly preferred over model 3b ($>3\sigma$), indicating that the flexibility that the intrinsically rocky population offers is better supported than adding additional flexibility to the intrinsically gaseous population. While we find that model 4b is somewhat preferred over model 4 ($<3\sigma$), we note that the fraction of intrinsically rocky planets $Q_\text{fr}$ increases rather than decrease or let alone become consistent with zero. While we cannot definitively prove the existence of intrinsically rocky planets, we have shown that their additions to the mixture model in the paper are justified. 

\subsection{Caveats}\label{Caveats}

While this work represents a major step forward in determining the occurrence rates of planets across a wide range of periods, masses and radii, and the interconnectedness of these properties, there exist several limiting factors that need to be taken into account when considering the results presented herein. We describe three broad caveats here, and continue the discussion in our mention of future work in Section 6.4.

\subsubsection{High dimensional distribution}\label{High dimensional distribution}

Traditional occurrence rates using \textit{Kepler} have naturally limited their scope to the radius-period plane, as these are the two quantities directly measured with the transit technique. Here, we have included masses of \textit{Kepler} planets measured with RV observations in order to expand this 2D distribution into a 3D distribution of mass, radius and period. In doing so, we have created models with up to 25 parameters and included three populations of planets to represent this complex 3D distribution. Despite this, the distribution presented here is still a projection of an even higher dimensional distribution. 

Mass, radius and period are the three main observables and some of the most fundamental quantities for exoplanets, but they just represent three of many possible factors that affect which planets are common, and how common those planets are. Together, planet radius and mass can be used as a proxy for composition, but it remains a proxy, and ultimately planet occurrence as a function of planet composition remains an important question. 

While here we have treated planets as independent, many \textit{Kepler} planets are found in multiple-planet systems, with these planets exhibiting uniformity in mass and radius, as well as regularly spaced orbits \citep{MillhollandEt2017, WeissEt2018AJ}. 

There are also a variety of factors associated with the host star that can affect planet occurrence. Host star metallicity has shown to have a significant effect not only on gas giant occurrence rates, but also sub-Neptune and terrestrial planets, with higher metallicities associated with higher planet occurrence \citep{Wang&Fischer2015AJ, MuldersEt2016AJ, DongEt2018PNAS, PetiguraEt2018AJ}. The presence of stellar companions has been linked to lower occurrence rates \citep{WangEt2014ApJ,  KrausEt2016AJ}. Stellar obliquity (alignment of planet orbits with respect to the spin of the host star)) is correlated with both planet size and period \citep{Munoz&Perets2018AJ}. While we only consider planets around G and K dwarfs in this work, planets around M dwarfs have been shown to have a different radius and mass distribution from those around G and K dwarfs, with a significant lack of gas giants \citep{BonfilsEt2013AAP, Dressing&Charbonneau2015ApJ}, and possible period-dependent differences \citep{MuldersEt2015bApJ, Neil&Rogers2018ApJ}.

Most planets discovered via transits and RV are relatively close to the Sun (within $\sim 1500$ parsecs), but the galactic distribution of planets could be substantially different, which will be measured with the future \textit{WFIRST} survey \citep{MontetEtAl2017PASP}. These factors, among others, could all affect planet occurrence, and the true distribution of planets has many dimensions.

\subsubsection{Data Set}\label{Data Set}

\textit{Kepler}'s discovery and characterization of several thousand exoplanets, and in particular the high-quality sample of 1130 planets used in this paper, has enabled the statistical study of the occurrence rate of planets in the 3D space of radius, period and mass. However, the statistical analysis will always be limited by the quality and quantity of the available data, and this paper is no exception. \textit{Kepler}'s detection efficiency drops off towards lower radii and longer periods, and these regions of high incompleteness are of high interest and contain many mysteries. The differences between the two separate populations of rocky planets modeled in this paper would manifest in these regions of high incompleteness where we do not have many planets in the sample. Extending the sample to longer orbital periods would allow us to more easily study the interplay between the period and radius/mass distributions, as the differences would be more pronounced.

The availability of mass measurements is an additional major limitation. Our main sample contains only 53 planets with RV mass measurements, yet we constrain the mass-radius relation for gaseous planets based on these 53 planets. Furthermore, there is a significant lack of RV measurements for small, potentially rocky planets. As a result, our rocky planet mass-radius relationship is fixed and based on theoretical curves: our planet sample would not be able to constrain this well at all. In order to model a population of gaseous planets together with a population of rocky planets, we cannot solely rely on the data, but must include physics in the form of hydrodynamic mass loss. With more RV mass measurements of planets in the size regime where there is some overlap between rocky and gaseous compositions (broadly $1.4 < \frac{R}{R_\oplus} < 2.0$), we would be able to more strongly constrain the difference in occurrence rates between the two populations. With the ongoing \textit{TESS} search for transiting planets and subsequent RV follow-up, we can expect this dataset to continue to grow in the future.

When compiling our dataset we made the choice of using RV mass measurements only, thereby excluding a comparable number of TTV masses. Planets with masses measured with RV vs TTV have been shown to have systematic differences in their densities. This is likely due to observational biases, with the TTV technique having higher sensitivity than RV to planets with lower masses and longer periods \citep{JontofHutterEt2014ApJ, Steffen2016MNRAS, Mills2017ApJL}. We exclude TTV planets in order to keep the dataset as homogeneous as possible, although we acknowledge that modeling the RV and TTV masses together is a critical next step. Given the differences in densities, we expect one major effect of including TTV masses would be to increase the scatter in the mass-radius relation. Further, since the TTV technique is more sensitive to longer period planets, including them may require extending our model to longer orbital periods than our analysis here. This may also allow further exploration of the connection between the mass-radius plane and orbital period, as mass-radius occurrence may depend on orbital period beyond the mixture models and envelope mass loss utilized in this paper. We leave more detailed analysis of the TTV sample to future work.

In our analysis, we have taken careful steps to fully account for the incompleteness of the \textit{Kepler} pipeline and subsequent dispositioning of planet candidates by the Robovetter software. While this should give us unbiased results as a function of radius and period, the incompleteness of RV followup of transiting planets has not been accounted for in our analysis and could bias our results. The RV followup process is less systematic and thus more difficult to properly account for. \citet{BurtEt18AJ} simulated several strategies for RV follow-up of \textit{TESS} planets, and found that the practice of only reporting masses measured with a high statistical significance lead to biased mass-radius relations. This bias results in a mass-radius relation pushed towards higher densities below $10 M_\oplus$, and has been found to affect previous empirical fits of the mass-radius relation. In our work, this could result in an underestimation of the number of low-mass gaseous planets, and a mass-radius relation that is shifted towards higher densities. In order to fully account for this RV selection effect, all non-detections regardless of statistical significance need to be reported (e.g. as done in \citet{MarcyEt2014ApJS}) and subsequently included in the data sample for papers doing statistical analysis.

\subsubsection{Model Dependence}\label{Model Dependence}

In this work we presented four models, of increasing complexity. As shown by our occurrence rate calculations in Section 5.2 and mass predictions in Section 5.4, our conclusions are highly dependent on the choice of model. For a planet of radius $1.9 R_\oplus$ and period 0.74 days, using model 1, we predict a mass distribution that peaks at $2 M_\oplus$, whereas we predict a mass distribution that peaks at $6-8 M_\oplus$ for the same planet using models 2-4. While we have put a lot of consideration into our choice of parametrizations for the models presented here, constructing them using both physical reasoning and past precedence, there remain many areas of potential improvement. For instance, in our model there is no constraint restricting the density of gaseous planets to be lower than that of a rocky planet of the same mass. As a result, high-density ``gaseous" planets are included in the calculation of $\eta_\oplus$. Changing the parametrization of one distribution (i.e. the period distribution) can have rippling effects throughout the entire 3D distribution. It is of the utmost importance to consider the results presented here in the context of the model that was used to generate them. 

Just as the choice of parametrization significantly affects the results, the choice of prior can often have a large effect. For instance, adopting a wide prior on $\alpha$, the parameter that governs the scaling for the envelope mass loss, can cause $\alpha$ to be retrieved several orders of magnitude higher than our default assumption of $\alpha = 1$, which is based on the default values for mass loss efficiency, stellar XUV flux, and age of the star assumed in \citet{Lopez&Fortney2013ApJ}. This would lead to planets that are more easily able to retain their envelopes, greatly reducing the number of evaporated cores. However, such high or low values of $\alpha$ may be unphysical ($\alpha<0.1$ or $\alpha> 10$) and would require the XUV flux of stars to strongly deviate from what we assumed. With this case and with other parameters, we have tried to strike a balance between restricting the prior to reasonable ranges of the parameter (based on physical models and past measurements), and allowing the model flexibility. Uncertainty ranges quoted herein represent credible intervals inferred in the context of a specified model. They do not include uncertainties in the choice of model parametrization or priors.

While our results are model dependent, our application of mixture models has shown to be justified by the available data, and we have shown that doing so requires the addition of the physics of envelope mass loss. We provide the results from multiple models, allowing us to assess how much this model choice impacts our conclusions. Finally, model dependence is not limited to our work and must be considered for any analysis of the mass, period, and/or radius distributions of exoplanets.

\subsection{How to use these results}\label{How to use these results}

The main purpose of this paper is to provide a framework for constraining higher dimensional planet occurrence rate distributions and including mixture models in these distributions. While we present occurrence rate measurements, mass/radius predictions, and measurements of the amount of gaseous vs evaporated core vs intrinsically rocky planets, our goal is not to put forward our results as definitive. All results must be considered in context of the model. Instead of ``we predict a mass of x", it is instead more prudent to say ``we predict a mass of x, if we use this model that includes hydrodynamic envelope mass loss". Broadly, when using our population inferences, we recommend using model 4, as this model includes the three populations of planets that are shown to be justified by the data in Section 5.3, and provides the best fit to the observed distribution of Kepler radii and periods. By providing fits to multiple models, we offer the exoplanet community the chance to test for robustness against the choice of model.

\subsection{Composition Distribution}\label{Composition Distribution}

Previous efforts to empirically constrain the radius-period distribution and mass-radius relationship of planets have relied on simple functions and distributions such as power-laws, normals and log-normals. While the physics of planet formation sometimes entered the scene, as in \citet{WolfgangEt2016ApJ} where planets are constrained to have a density lower than that of a pure iron core, these scenarios are few and far between and there remains a gulf between observations and theory. Here, we have included the physics of envelope mass loss as a necessary step to model rocky planets alongside gaseous planets and disfavor unphysical small planets with gaseous envelopes on short orbits. This can make the model somewhat incongruous and unsatisfying: for some aspects we rely on physics, and for some we trust the data to constrain the distributions for us.

One possible step towards a more physically grounded model would be to constrain a composition distribution. Rather than working with the observables mass and radius, we can choose mass and some proxy for composition, such as envelope mass fraction. Using input from theoretical models of planet interior structure, we can go from mass and envelope mass fraction to an output radius. Thus, we should be able to recreate the double broken power-law mass-radius relationship described here, but as a natural outcome of planet interior structure. As we obtain more and more planet mass measurements, and reveal areas in the planet mass-radius space where composition visibly differs, moving towards this composition distribution will become easier and more necessary. With such a distribution, we can close the gap between theory and observations in our modelling.

\section{Conclusion}\label{Conclusion}

We have used hierarchical Bayesian modeling to constrain the joint mass-radius-period distribution of \textit{Kepler} transiting exoplanets. We construct four models of increasing complexity, the latter three using mixture models and XUV-driven hydrodynamic mass loss to independently model populations of planets with gaseous envelopes, evaporated cores, and intrinsically rocky planets. We find through model selection techniques that these additional complexities are warranted, and better match the observed radius and period distributions of \textit{Kepler}. We calculate occurrence rates in different regions of parameter space using these models, including extrapolation to $\eta_\oplus$, finding broad agreement with previous results but also finding the estimates of $\eta_\oplus$ to be significantly model dependent. We use these models to predict masses of \textit{Kepler} planets, showing the benefit of employing mixture models by comparing the predictions for two planets of similar radius but dissimilar periods. Although our results are highly dependent on the parametrization and choice of priors for our models, we provide a framework for the modelling of higher-dimensional distributions of exoplanets together with the inclusion of mixture models. We find the use mixture models to be a valuable tool for capturing the complexity inherent to planet formation and evolution.

\bigskip

We thank our anonymous referee for providing valuable feedback and suggestions that improved the paper. LAR gratefully acknowledges support from NASA Exoplanet Research Program grant NNX15AE21G and from NSF FY2016 AAG Solicitation 12-589 award number 1615089.

\bibliography{exoplanets}

\end{document}